\documentclass{article}
\usepackage{amsmath,amssymb}
\usepackage[hyper=false]{pik}
\usepackage[authoryear]{natbib}
\usepackage{bm}
\usepackage{rotating}
\usepackage{xcolor}
\usepackage{multirow}

\title{Comparison of electricity market designs for stable decentralized power grids}
\author{Jobst Heitzig, Potsdam Institute for Climate Impact Research, PO Box 60 12 03, 14412 Potsdam, Germany
	\and Hildegard Meyer-Ortmanns, Jacobs University, Campus Ring 1, 28759 Bremen, Germany
	\and Sabine Auer, Potsdam Institute for Climate Impact Research, PO Box 60 12 03, 14412 Potsdam, Germany
}
\date{~\\~\\\today}

\def\rho{\varrho}
\def\ge{\geqslant}

\def\le{\leqslant}

\def\eps{\varepsilon}

\begin{document}

\maketitle

\section{Introduction}

In this study,
we develop a theoretical model of strategic equilibrium bidding and price-setting behaviour 
by heterogeneous and boundedly rational electricity producers and a grid operator in a single electricity market
under uncertain information about production capabilities and electricity demand.

We compare eight different market design variants 
and several levels of centralized electricity production 
that influence the spatial distribution of producers in the grid,
their unit production and curtailment costs,
and the mean and standard deviation of their production capabilities.

Our market design variants differ in three aspects.
Producers are either paid their individual bid price (``pay as bid'') 
or the (higher) market price set by the grid operator (``uniform pricing'').
They are either paid for their bid quantity (``pay requested'')
or for their actual supply (``pay supplied'') which may differ due to production uncertainty.
Finally, excess production is either required to be curtailed or may be supplied to the grid.

After formally deriving producers' and grid operator's optimization problems
and finding their analytical intractability, 
we then use numerical simulations to study their resulting strategic behaviour
in an agent-based model slightly similar to \cite{mureddu2015green} with 200 producers on 100 grid nodes 
for 50 repetitions of each of the eight markets and ten different synthetically generated network topologies.

We analyse results w.r.t.\ several aggregate indicators such as
the average bid price-to-cost ratio, market price, bid quantity-to-mean production ratio,
balancing needs, total economic costs, producers' profits, consumers' costs,
and total grid workload.

Overall, we find the combination of uniform pricing, paying for requested amounts, and required curtailment
to perform best or second best in many respects, and to provide the best compromise between the goals
of low economic costs, low consumer costs, positive profits, low balancing, low workload, and honest bidding behaviour.

\section{Model}

Our model simulates the bidding and price-setting decisions 
of a number of heterogeneous electricity producers and their power grid operator
under eight different generalized market designs
and different degrees of (de-)centralized production.
We focus on the market for a specific time-slot $T$,
e.g., a certain 15-minute interval of the day,
and assume that decisions have to be taken at an earlier time $t < T$,
so that certain quantities such as demand and actual production at time $T$
are uncertain at time $t$ and have to be described via random variables.

\begin{figure}
\begin{center}
\includegraphics[width=\textwidth]{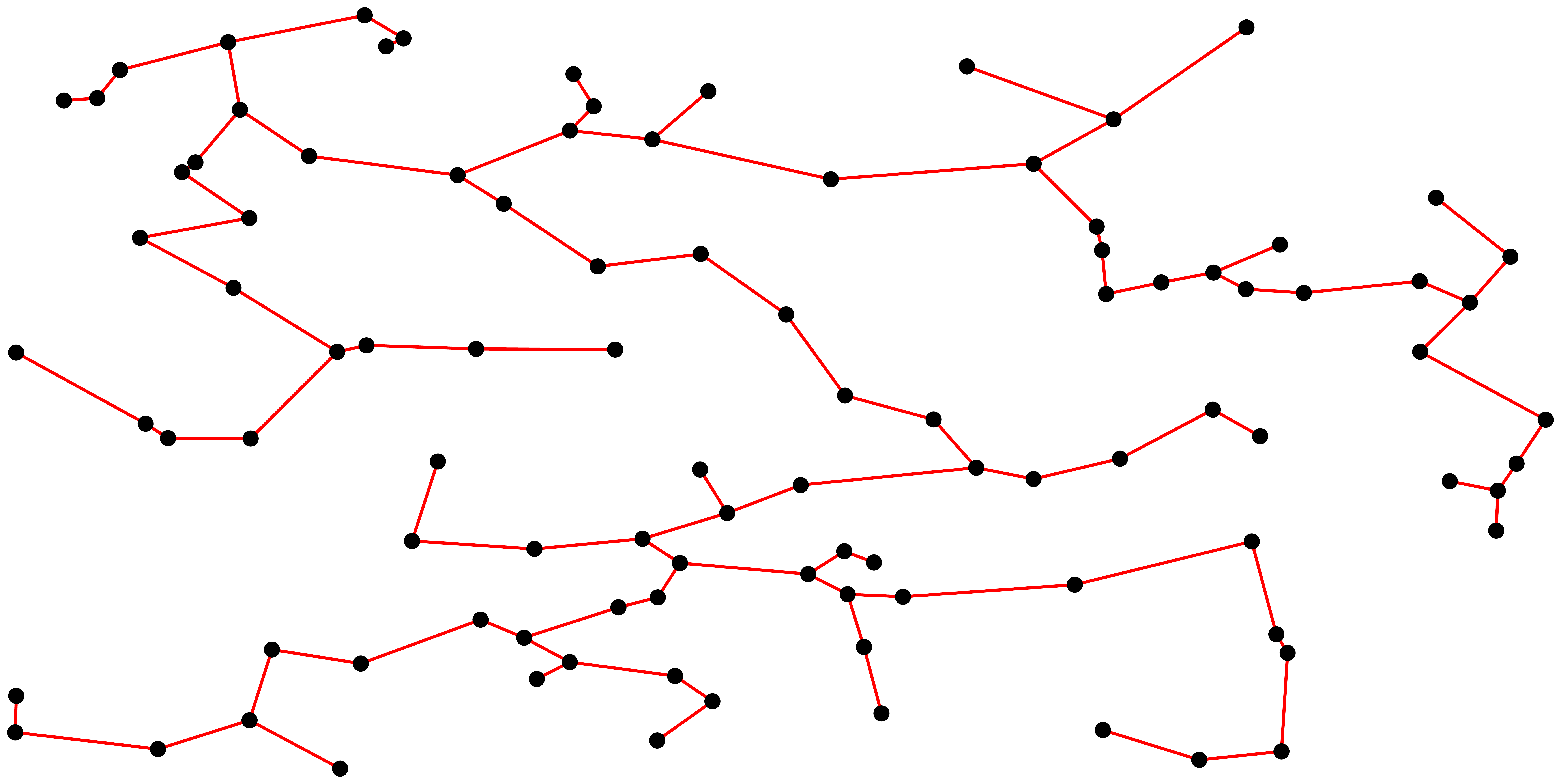}
\end{center}
\caption{\label{fig:grid} Sample grid generated according to \cite{schultz2014random}.}
\end{figure}

\paragraph{Power grid.}
We use an ensemble of ten different {\em single-level power grids} consisting of $n=100$ nodes each
placed at positions chosen uniformly at random from a square territory
and connected via the synthetic random power-grid generation algorithm developed earlier in \cite{schultz2014random}.
To simplify power flow computations, we assume the grid to be tree-shaped and consist of lossless lines.
See Fig.\,\ref{fig:grid} for an example grid.

\paragraph{Electricity demand.}
At each node $v_a$, there will be an electricity {\em demand} at time $T$
whose uncertain value is described at time $t$ by a random variable $d_a$.
We assume $d_a$ to be independently normally distributed 
with mean $\mu^d_a$ and standard deviation $\sigma^d_a = \mu^d_a / 4$,
where $\mu^d_a$ is larger for more decentral nodes, representing the fact that demand is rather decentralized.
We measure a node $v$'s centrality by the standard metric of {\em closeness centrality} from complex network theory,
which is given by $CC(v) = 1 / \sum_{v'} d(v,v')$, 
where $v'$ runs over all nodes and $d(v,v')$ is the number of transmission lines one has to pass to get from $v$ to $v'$.
We enumerate all nodes by ascending closeness centrality, 
so that $CC(v_1)\le CC(v_2)\le \ldots \le CC(v_n)$,
and put $\mu^d_a = (n + 1 - a) / n (n + 1)$.
{\em Total demand} $D$ thus has mean $\mu^D = \sum_a \mu^d_a = 1 / 2$ 
and standard deviation $\sigma^D = \sqrt{\sum_a (\sigma^d_a)^2} \approx 1 / \sqrt{3 n} \approx .057$,
i.e., with a relative variability of about 11\%.

\paragraph{Production capabilities.}
Electricity is produced by $N=200$ independent and heterogeneous {\em producers}
which range from large-scale utilities to smallest-scale producers such as single households with a number of PV panels.
The spatial distribution of producers onto grid nodes is governed by a
{\em centralization parameter} $-1 \le  Z \le 1$, 
where $ Z = 1$ generates a rather centralized production structure dominated by traditional large-scale utilities,
while $ Z = -1$ generates a rather decentralized production structure dominated by variable renewable energy sources (VRES).
Depending on $ Z$, producers are placed onto nodes at random,
either favouring more central nodes (for positive $ Z$) or more decentral nodes (for negative $ Z$).
More precisely, each producer $i = 1\ldots N$ is then placed at a random node $v(i)$,
where $v(i) = v_a$ with probability $(1 +  Z (2a - n - 1) / n) / n$.

Each producer $i$ will either produce or not produce electricity at time $T$.
If they produce at time $T$, they will produce an amount of electricity 
that is only known with some uncertainty at time $t$,
so {\em production} is described by a random variable $u_i$.
To simplify calculations, we assume $u_i$ to be 
uniformly distributed between two known limits $\mu_i (1 - \rho_i)$ and $\mu_i (1 + \rho_i)$.
To represent the fact that decentralized production tends to by smaller-scale and more variable,
we assume that mean production $\mu_i$ is positively correlated with the centrality of $i$
and $\rho_i$ is negatively correlated with centrality,
by putting $\mu_i = a \mu_0 / n$ and $\rho_i = (n + 1 - a) \rho_0 / n$,
where $a$ is the node index of $v(i) = v_a$.
We chose $\mu_0$ so that the {\em mean total production potential} $\sum_i \mu_i = 1$, 
i.e., equals twice the mean demand $\mu^D = 1 / 2$
to get a considerable but fixed amount of {\em competition} between producers.
We put $\rho_0 = 1/2$ so that the most decentral nodes will have an uncertain production
ranging from $\mu_i / 2$ to $3\mu_i / 2$, while the most central nodes have an almost certain production.

Note that for simplicity, we ignore the actually existing but nontrivial spatial correlation 
in both demand and production in this study.

\paragraph{Production, balancing, and curtailment costs.}
Producer $i$'s {\em unit production costs} $c_i$ are assumed to be known for certain and also depend on centrality,
$c_i = a / n$, where $a$ is again the node index of $v(i) = v_a$,
representing the fact that renewable energy sources have lower production costs.
Hence {\em expected mean unit production costs} are $\bar c \approx (3 +  Z) / 6 \in [1/3, 2/3]$, 
which will serve us as a reference point for the emerging market price later
and represents a potential production cost reduction of 50\% 
when going from rather centralized production to rather decentralized production.

If at time $T$, actual power supply falls short of demand, 
the grid operator needs to buy upward balancing energy.
Since we do not want to study interactions between different markets such as spot and balancing market here
but focus on the design of one market stage only, 
we assume here that the {\em unit upward balancing costs} $c^+$ are fixed and known already at time $t$,
and we put $c^+ = 2$, i.e., between three and six times the mean unit production costs $\bar c$, 
depending on the degree of centralization $ Z$.
Analogously, if supply exceeds demand, the grid operator has to buy some downward balancing demand
at a fixed price $c^-$. 
We choose a relatively large value of $c^- = 1/4$ 
to represent the assumption that in view of the need to reduce energy demand,
society must give the grid operator a considerable incentive to avoid excess supply.

In some of our market designs, if producer $i$ produces more than agreed with the grid operator,
$i$ is not allowed to supply this excess production to the grid but is required to ``curtail'' it.
If this is the case, we assume $i$'s {\em unit curtailment costs} are $c^0_i = c^0_0 (n + 1 - j) / n$,
representing the fact that smaller, more decentralized producers tend to have larger curtailment costs.
On average, however, we assume that curtailment is rather cheap, being only one percent of mean production costs,
$c^0_0 = c_0 / 100$.

\paragraph{Penalties.}
In all of our market designs, we assume that producers 
actually supplying more or less electricity to the grid than agreed with the grid operator
have to pay some penalties to the latter.
These penalties serve a double purpose,
they provide an incentive for the producers to make realistic quantity bids
and they compensate the grid operator for balancing costs caused by supply excesses or shortfalls.
Because of the latter,
we assume that for each unit of electricity supplied more than agreed one has to pay 
a {\em unit excess supply penalty} $c^>$ that equals downward balancing cost, $c^> = c^-$,
while for each unit supplied less than agreed one has to pay
a {\em unit supply shortfall penalty} $c^<$ that equals upward balancing cost, $c^< = c^+$.
Note that penalties have to be paid regardless of whether at time $T$ there is actually any need for balancing.
One may consider alternative market designs in which penalties reflect actual instead of potential balancing costs,
but it turns out that finding optimal bids in such a scenario is much more complicated than in our setting,
so we leave this possibility for future research.

\paragraph{Overall market procedure.}
At time $t$, each producer $i$ places a {\em price-and-quantity bid} $(p_i,q_i)$,
thereby stating that they offer to supply $q_i\ge 0$ units of power at time $T$ for a unit price of $p_i\ge 0$.
They are free to choose any combination of $p_i,q_i$,
hence {\em bid price} $p_i$ is not required to be related in any way to unit production costs $c_i$
and {\em bid quantity} $q_i$ is not required to be related in any way to mean production $\mu_i$.
However, as we assume producers to be boundedly rational 
and will chose a combination of $p_i,q_i$ that seems economically optimal for them,
i.e., for which they expect their profit to be maximal, 
in view of their past observation of the emerging market price
and the information available at time $t$ about the uncertain demand and production capabilities.
We will see below that in general, $p_i > c_i$, and typically $\mu_i (1-\rho_i) < q_i < \mu_i$.

After all bids are made, the grid operator selects a set $A$ of producers whose bids they accept,
trying to minimize their expected total costs in view of the uncertain information available at time $t$.
To reduce the complexity of this optimization problem,
we assume that the grid operator does not consider all of the $2^N$ many possible sets $A$ of producers
but only those sets $A$ that result from choosing a {\em cutoff price} $P\ge 0$
and accepting all cheaper bids, i.e., putting $A = \{ i:p_i \le P \}$.
This is motivated by the fact that if production capabilities and demand were certain,
the cheapest way to meet demand $D$ would be to place the bids into a ``merit order''
so that $p_1\le p_2\le \ldots\le p_n$
and then choose $P$ so that $\sum_{i:p_i\le P} q_i\approx D$.
In princinple, when $D$ and $u_i$ are uncertain, 
such a scheme for choosing $A$ might not actually be optimal 
since in view of the high potential costs of balancing mismatches,
it may sometimes be preferable to accept a more expensive but more certain bid (e.g.\ by a traditional power plant)
rather than a cheaper but more uncertain bid (e.g.\ by a small PV installation).
For two reasons, 
we ignore this fact here and still assume that the grid operator accepts all bids below a certain cutoff price.
For one, the more general optimization problem would require much more information about production uncertainties
and would be much harder to solve in reality, so that using the simpler scheme can be interpreted as some form of bounded rationality.
Second, using the more general scheme would probably result in a bias towards traditional electricity producers
which we do not want to assume here as it would contradict the observable political tendency to rather favour renewable production,
so we choose a scheme that is more unbiased between the two forms.

After the cutoff price $P$ was set at time $t$,
all producers with $p_i\le P$ will produce some actual amount of electricity $u_i$ at time $T$,
leading to production costs $c_i u_i$,
may have to curtail the excess production $e_i = \max(u_i - q_i, 0)$ (in some marked designs),
leading to curtailment costs $c^0_i e_i$
They then {\em supply} some amount $s_i$ to the grid,
where either $s_i = \min(u_i, q_i) = u_i - e_i$ or $s_i = u_i$, 
depending on whether curtailment was required or not.
Since supply may still deviate from the requested amount of electricity that is given by the accepted bid quantity, $r_i = q_i$,
this may lead to either a positive amount of {\em excess supply} $s^+_i = \max(s_i - r_i, 0)$
or to a positive amount of {\em supply shortfall} $s^-_i = \max(r_i - s_i, 0)$,
and thus to total penalties $c^+ s^+_i + c^- s^-_i$ they pay to the grid operator.
The grid operator, on the other hand, pays each producer with $p_i\le P$ 
either for the requested amount $r_i$ (``pay requested'') 
or for the actually supplied amount $s_i$ (``pay supplied''),
and either at a uniform price $P$ (``uniform pricing'')
or at each producer only at their bid price $p_i \le P$ (``pay as bid''),
depending on the market design variant.

\paragraph{Market design variants.}
The option of either requiring curtailment (``curtail'') or not (``don't curtail''),
paying for either requested or supplied amounts, and paying either a uniform price or as bid,
we thus get $2^3 = 8$ different market design variants to compare,
each of which sets different incentives for producers and grid operator
and will thus imply different optimizing behaviour regarding the choice of $(p_i,q_i)$ and $P$,
leading to different strategic equilibria and finally to different 
total economic costs of the system, total producers' profits, 
total curtailment and balancing, total workload on the grid, etc.

\paragraph{Experiment design.}
We assume the market is run repeatedly under similar conditions, 
e.g., each day of similar weather conditions and for the same time of day $T$,
so that producers can form expectations about the market price $P$
that they can incorporate into their optimization 
by treating $P$ as a random variable whose distribution matches the past distribution of observed values of $P$.
Since this leads to a relatively fast convergence of bidding behaviour over time,
we run each market variant 50 times in sequence for each of the ten power grids from the ensemble,
resulting in a sample of 500 observations per market variant,
so that statistical averages over this sample will have a standard error of only 
$\sqrt{1/500}\approx 4.5\%$ of their respective standard deviation,
which we will use in testing statistically whether market variants differ significantly 
with respect to the above-mentioned indicators.

\begin{table}\begin{center}\begin{tabular}{ll}\toprule
 $-1 \le  Z \le 1$ & degree of centralization \\
 $d_a$ & uncertain demand at grid node $v_a$ \\
 $D = \sum_a D_a$ & uncertain total demand \\
 $u_i \ge 0$ & uncertain production of producer $i$ \\
 $c_i > 0$ & known unit costs of production \\
 $c^+ > 0$ & known price of upward balancing by grid operator \\
 $c^- > 0$ & known price of downward balancing by grid operator \\
 $c^<$ & known penalty for supplying one unit less than requested \\
 $c^>$ & known penalty for supplying one unit more than requested \\
 $c^0_i > 0$ & known costs of curtailment \\
 $p_i,q_i \ge 0$ & producer's strategic bid price and quantity \\
 $P\ge 0$ & grid operators strategically chosen cutoff price \\
 $r_i \in \{0, q_i\}$ & resulting quantity requested from $i$ \\
 $s_i, S = \sum_i s_i \ge 0$ & actual producer's and total supply \\
 $s_i^< = \max(r_i - s_i, 0) \ge 0$ & producer's supply shortfall \\
 $s_i^> = \max(s_i - r_i, 0) \ge 0$ & producer's excess supply \\
 $S^< = \sum_i s_i^< \ge 0$ & total supply shortfalls \\
 $S^> = \sum_i s_i^> \ge 0$ & total excess supplies \\
 $B^+ = \max(D - S, 0) \ge 0$ & actual upward balancing needs \\
 $B^- = \max(S - D, 0) \ge 0$ & actual downward balancing needs\\\bottomrule
\end{tabular}\end{center}\caption{\label{tbl:notation}Overview of symbols}\end{table}

\subsection{Producers' and grid operator's behaviour}
The central ingredient of the model is our assumption of how producers and grid operator 
make their choice of $(p_i,q_i)$ and $P$.

\paragraph{Producers.}
Let us consider the optimization problem of a fixed producer $i$,
and thus drop the index $i$ from all quantities for now.
In all market variants, $i$'s profit $\pi$ is either zero (if $P < p$) 
or of the general form
\begin{align}
	\pi &= \beta q + \gamma p q + \delta P q + \eps u + \zeta p u + \eta P u
\end{align}
(if $P \ge p$ and $u \le q$) or
\begin{align}
	\pi &= \beta' q + \gamma' p q + \delta' P q + \eps' u + \zeta' p u + \eta' P u
\end{align}
(if $P \ge p$ and $u > q$),
with coefficients $\beta,\beta',\ldots,\eta,\eta'$ that depend on the market design variant,
see Table~\ref{tbl:coeffs}.

\begin{table}\begin{center}\begin{tabular}{lllllllll}\toprule
 & \multicolumn{4}{l}{uniform pricing}               & \multicolumn{4}{l}{pay as bid} \\
 & \multicolumn{2}{l}{pay supplied} 
 						   & \multicolumn{2}{l}{pay requested} 
 						                             & \multicolumn{2}{l}{pay supplied} 
 						                                                       & \multicolumn{2}{l}{pay requested} \\
 & don't curtail & curtail & don't curtail & curtail & don't curtail & curtail & don't curtail & curtail \\\midrule
$\beta'$  & $c^> $ & $ $ & $c^> $ & $ $ & $c^> $ & $ $ & $c^> $ & $ $ \\ 
$\gamma$  & $ $ & $ $ & $ $ & $ $ & $ $ & $ $ & $1 $ & $1 $ \\ 
$\gamma'$ & $ $ & $ $ & $ $ & $ $ & $ $ & $1 $ & $1 $ & $1 $ \\ 
$\delta$  & $ $ & $ $ & $1 $ & $1 $ & $ $ & $ $ & $ $ & $ $ \\ 
$\delta'$ & $ $ & $1 $ & $1 $ & $1 $ & $ $ & $ $ & $ $ & $ $ \\ 
$\eps'$   & $ -c-c^>$ & $ -c-c^0$ & $ -c-c^>$ & $ -c-c^0$ & $ -c-c^>$ & $ -c-c^0$ & $ -c-c^>$ & $ -c-c^0$ \\ 
$\zeta$   & $ $ & $ $ & $ $ & $ $ & $1 $ & $1 $ & $ $ & $ $ \\ 
$\zeta'$  & $ $ & $ $ & $ $ & $ $ & $1 $ & $ $ & $ $ & $ $ \\ 
$\eta$    & $1 $ & $1 $ & $ $ & $ $ & $ $ & $ $ & $ $ & $ $ \\ 
$\eta'$   & $1 $ & $ $ & $ $ & $ $ & $ $ & $ $ & $ $ & $ $ \\
$\theta'$   & $ -c^>$ & $ $ & $ -c^>$ & $ $ & $ -c^>$ & $ $ & $ -c^>$ & $ $ \\\bottomrule
\end{tabular}\end{center}
\caption{\label{tbl:coeffs}Nonzero optimization coefficients by market design variant.
Always $\beta = -c^<$, $\eps = c^< - c$, and $\theta = c^<$.}
\end{table}

The random variables $P$ (assumed to be distributed as observed in the past) and $u$
are independent, hence expected profit given some choice of $(p_i,q_i)$ is
\begin{align}
	E \pi &= \int_p^\infty dP\,\phi_P(P)\left[
		\int_0^q du\,\phi_u(u) [ \beta q + \gamma p q + \delta P q + \eps u + \zeta p u + \eta P u ] \right. \nonumber \\
	&\quad\quad\quad\quad\quad\quad\quad\left. 
		+ \int_q^\infty du\,\phi_u(u) [ \beta' q + \gamma' p q + \delta' P q + \eps' u + \zeta' p u + \eta' P u ]
	\right] \\
	&= \int_p^\infty dP\,\phi_P(P)\left[
		\Phi_u^<(q) [ \beta q + \gamma p q + \delta P q ] 
		+ E_u^<(q) [ \eps + \zeta p + \eta P ] \right. \nonumber\\
	&\left.\quad\quad\quad\quad\quad\quad\quad{}
		+ \Phi_u^>(q) [ \beta' q + \gamma' p q + \delta' P q ]
		+ E_u^>(q) [ \eps' + \zeta' p + \eta' P ]
	\right] \\
	&= \Phi_P^>(p) \left[
		\Phi_u^<(q) [ \beta q + \gamma p q ] 
		+ E_u^<(q) [ \eps + \zeta p ] 
		+ \Phi_u^>(q) [ \beta' q + \gamma' p q ]
		+ E_u^>(q) [ \eps' + \zeta' p ]
	\right] \nonumber \\
	&\quad
	+ E_P^>(p) \left[
		\Phi_u^<(q) \delta q 
		+ E_u^<(q)\eta
		+ \Phi_u^>(q) \delta' q 
		+ E_u^>(q) \eta'
	\right],
\end{align}
where
$\phi_x(\cdot), \Phi^<_x(\cdot)$ and $\Phi^>_x(\cdot) = 1 - \Phi^<_x(\cdot)$
are the probability density, cumulative and decumulative distribution functions 
of random variable $x$,
and we introduced the abbreviations 
\begin{align}
	E_x^<(y) &= \int_0^y dx\,\phi_x(x) x,\\
	E_x^>(y) &= \int_y^\infty dx\,\phi_x(x) x.
\end{align}
For a maximum of $E\pi$ with $p,q>0$, these first-order conditions must be met:
\begin{align}
	0 = \partial_p E \pi &= - \phi_P(p)\left[
		\Phi_u^<(q) [ \alpha + \beta q + (\gamma + \delta) p q ] 
		+ E_u^<(q) [ \eps + (\zeta + \eta) p ] \right. \\
	&\quad\quad\quad\quad\left.
		+ \Phi_u^>(q) [ \alpha' + \beta' q + (\gamma' + \delta') p q ]
		+ E_u^>(q) [ \eps' + (\zeta' + \eta') p ] ]		
	\right] \\
	&\quad\quad
	+ \Phi_P^>(p) \left[
		\Phi_u^<(q) \gamma q 
		+ E_u^<(q) \zeta 
		+ \Phi_u^>(q) \gamma' q 
		+ E_u^>(q) \zeta'
	\right], \\ 
	0 = \partial_q E \pi &= \Phi_P^>(p) \left[
		\phi_u(q) [ \Delta\alpha + \Delta\beta q + \Delta\gamma p q ]
		+ \phi_u(q) q [ \Delta\eps + \Delta\zeta p ] \right. \\
	&\quad\quad\quad\quad\left.
		+ \Phi_u^<(q) [ \beta + \gamma p ] 
		+ \Phi_u^>(q) [ \beta' + \gamma' p ] 
	\right] \\
	&\quad\quad
	+ E_P^>(p) \left[
		\phi_u(q) [\Delta\delta q + \Delta\eta]
		+ \Phi_u^<(q) \delta 
		+ \Phi_u^>(q) \delta'
	\right],
\end{align}
where $\partial_p = \frac{\partial}{\partial p}$, $\Delta\alpha = \alpha - \alpha'$ etc.
Although this has no closed form solution,
it allows us to find $i$'s optimal values $p_i,q_i$ given the distributions $\phi_{u_i}$, $\phi_P$
numerically,
and it turns out that there generally $E\pi_i$ depends concavely on $p_i$ and $q_i$, 
hence the unique local maximum is the global maximum.
In general, there is an incentive to bid a low price to increase the chances of being selected,
but depending on the market variant,
there is a counteracting incentive to bid a higher price to make sure that the paid price is high enough to offset production costs and penalties.
Also, it may be optimal to bid a quantity lower than average production, to avoid costly curtailment or excess suppy penalty,
or to bid above average production to avoid penalty for shortfalls and/or get excess payments for oversupply.
If quantity bids are rather low, the merit order will be steeper, 
so the resulting $P$ will be larger to meet expected demand,
which will explain part of the observed price differences.

\paragraph{Grid operator.}
In view of all bids $(p_i,q_i)$ and the uncertain demand $D$ and production capabilities $u_i$,
the grid operator's expected profit given some choice of $P$ is
\begin{align}
	E\Pi &= \sum_{i:p_i\le P}\Big[			
		\Phi_u^<(q) (\beta_i q_i + \gamma_i p_i q_i) + E_u^<(q) (\theta_i + \zeta_i p_i)
        + P ( \Phi_u^<(q) \delta_i q_i + E_u^<(q) \eta_i ) \nonumber\\
    &\quad\quad\quad\quad
        + \Phi_u^>(q) (\beta'_i q_i + \gamma'_i p_i q_i) + E_u^>(q) (\theta'_i + \zeta'_i p_i)
        + P ( \Phi_u^>(q) \delta'_i q_i + E_u^>(q) \eta'_i )
		\Big] \nonumber\\
    &\quad
    	- c^+ EB^+ - c^- EB^-,		
\end{align}
where the expected balancing amounts $EB^+,EB^-$ are determined as follows.
Supply $s_i$ of producer $i$ with $p_i\le P$ has expected value $Es_i = \mu_i$
in the ``don't curtail'' market design variants, and
\begin{align}
	Es_i &= \left\{\begin{array}{ll}
		q_i   & \text{if~} q_i \le q^l_i \\
		\mu_i & \text{if~} q_i \ge q^h_i \\
		\frac{(q_i^2 - (q^l_i)^2) / 2 + (q^h_i - q_i) q_i}{q^h_i - q^l_i} & \text{otherwise} 
	\end{array}\right.
\end{align}
in the ``curtail'' market variants,
where $q^l_i = \mu_i (1 - \rho_i)$ and $q^u_i = \mu_i (1 + \rho_i)$.
The corresponding variances are
$Var(s_i) = (q^h_i - q^l_i)^2 / 12$ for ``don't curtail'', and
\begin{align}
	Var(s_i) &= \left\{\begin{array}{ll}
		0   & \text{if~} q_i \le q^l_i \\
		(q^h_i - q^l_i)^2 / 12 & \text{if~} q_i \ge q^h_i \\
		\frac{ (q_i - q^l_i)^3 / 12}{q^h_i - q^l_i} & \text{otherwise} 
	\end{array}\right.
\end{align}
in the ``curtail'' case.
As we assume production at different places to be independent and since $N$ is large,
total supply $S$ is thus approximately normally distributed
and its mean and variance can be computed as 
$ES = \sum_{i:p_i\le P}Es_i$ and $Var(S) = \sum_{i:p_i\le P}Var(s_i)$.
Since production capabilities are independent of demand,
excess demand $D - S$ is thus also approximately normally distributed
and its mean and variance can be computed as 
$m = E(D-S) = \mu^D - ES$ and $v = Var(D-S) = (\sigma^D)^2 + Var(S)$.
Finally, $EB^\pm$ is the expected value of $B^+ = \max(D - S, 0)$ and $B^- = \max(S - D, 0)$, 
given by
\begin{align}
	EB^\pm &= \exp(-m^2 / 2v) \sqrt{v/2\pi} \pm m [1 + \text{erf}(\pm m / \sqrt{2v})] / 2,
\end{align}
where erf is the error function.

\section{Results}

\begin{sidewaystable}[p]
\begin{footnotesize}
\begin{center}
\begin{tabular}{ll|rrrr|rrrr}\toprule
centrali- &  & \multicolumn{4}{c|}{uniform pricing} & \multicolumn{4}{c}{pay as bid} \\ 
zation  &  & \multicolumn{2}{c}{pay supplied} & \multicolumn{2}{c|}{pay requested} & \multicolumn{2}{c}{pay supplied} & \multicolumn{2}{c}{pay requested}   \\ 
$Z$ &  & \multicolumn{1}{l}{don't curtail} & \multicolumn{1}{r}{curtail} & \multicolumn{1}{l}{don't curtail} & \multicolumn{1}{r|}{curtail} & \multicolumn{1}{l}{don't curtail} & \multicolumn{1}{r}{curtail} & \multicolumn{1}{l}{don't curtail} & \multicolumn{1}{r}{curtail}  \\ \midrule
1 & price $P$ & \textbf{0.84064} & 1.25805 & 1.19578 & 1.20318 & 1.42060 & \color{red}\bf2.01057 & 1.61220 & 1.61784  \\ 
 & total economic costs & 0.61720 & 0.44271 & 0.44379 & \textbf{0.39156} & 0.58410 & 0.44417 & 0.42890 & \textbf{\textit{0.40245}}  \\ 
 & std. total economic costs & 0.03546 & 0.04000 & 0.03391 & 0.03247 & 0.03589 & 0.06420 & 0.03419 & 0.01820  \\ 
 & producers' profits $\pi$ & \color{red}\bf-0.00083 & 0.14278 & 0.15460 & \textbf{\textit{0.19179}} & 0.16811 & \color{red}\bf0.58631 & 0.33890 & \color{blue}{0.42652}  \\ 
 & std. producers' profits & 0.01166 & 0.02381 & 0.01400 & 0.01985 & 0.01617 & 0.03587 & 0.01450 & 0.01900  \\ 
 & grid workload & \textbf{1.04979} & \textbf{1.04126} & 1.09562 & \textit{1.06110} & 1.17799 & 1.27188 & 1.14883 & 1.16823  \\ 
 & std. grid workload & 0.15185 & 0.15531 & 0.09036 & 0.10222 & 0.16500 & 0.20746 & 0.13520 & 0.13990  \\ 
 & curtailment & 0.00000 & 0.02257 & 0.00000 & 0.02176 & 0.00000 & 0.03096 & 0.00000 & 0.02340  \\ 
 & upward balancing $B^+$ & \color{red}\bf0.23808 & 0.07222 & 0.08013 & 0.03093 & 0.18250 & 0.00040 & 0.05399 & 0.00251  \\ 
 & downward balancing $B^-$ & 0.00000 & 0.00011 & 0.00000 & 0.00045 & 0.00000 & 0.05574 & 0.00001 & 0.02365  \\ 
 & $(q_i-\mu_i)/\rho_i\mu_i$ & -0.30039 & -0.14183 & -0.00525 & -0.02410 & -0.29392 & -0.08736 & 0.07148 & 0.06535  \\ 
 & bid price-to-cost ratio $p_i/c_i$ & \textbf{1.89554} & 2.35714 & 2.43711 & \textbf{\textit{2.07482}} & 3.07774 & 3.90803 & 3.31576 & 3.22720  \\ \midrule
0.5 & price & \textbf{0.74170} & 1.19551 & 1.13596 & 1.13354 & 1.37618 & \color{red}\bf2.06409 & 1.57352 & 1.57957  \\ 
 & total economic costs& 0.63976 & 0.42089 & 0.42092 & \textbf{\textit{0.37836}} & 0.57897 & 0.46414 & 0.43300 & \textbf{0.37394}  \\ 
 & std. total economic costs& 0.03341 & 0.03956 & 0.03643 & 0.03416 & 0.03908 & 0.14286 & 0.03873 & 0.01800  \\ 
 & producers' profits & \color{red}\bf-0.02028 & 0.11986 & 0.13960 & \textbf{\textit{0.16936}} & 0.15698 & \color{red}\bf0.64996 & 0.31956 & \color{blue}{0.41058}  \\ 
 & std. producers' profits & 0.01017 & 0.02092 & 0.01548 & 0.01777 & 0.01672 & 0.18815 & 0.01607 & 0.01635  \\ 
 & grid workload & \textbf{0.93604} & 0.96808 & 1.02369 & \textbf{\textit{0.96542}} & 1.13220 & 1.25796 & 1.08555 & 1.05780  \\ 
 & std. grid workload & 0.15379 & 0.11895 & 0.17079 & 0.12828 & 0.17196 & 0.35750 & 0.14155 & 0.14310  \\ 
 & curtailment & 0.00000 & 0.02469 & 0.00000 & 0.02657 & 0.00000 & 0.04010 & 0.00000 & 0.02591  \\ 
 & upward balancing & \color{red}\bf0.26836 & 0.07175 & 0.08070 & 0.03873 & 0.19133 & 0.00038 & 0.06960 & 0.00386  \\ 
 & downward balancing & 0.00000 & 0.00026 & 0.00000 & 0.00006 & 0.00000 & 0.09052 & 0.00008 & 0.01537  \\ 
 & $(q_i-\mu_i)/\rho_i\mu_i$ & -0.26047 & -0.12842 & -0.00500 & -0.02639 & -0.25559 & -0.07131 & 0.06967 & 0.06123  \\ 
 & bid price-to-cost ratio $p_i/c_i$ & \textbf{2.21957} & 3.25872 & 4.02303 & \textbf{\textit{2.92172}} & 4.72686 & 6.39770 & 5.46417 & 5.11118  \\ \midrule
0 & price & \textbf{0.61963} & 1.09598 & 1.04492 & 1.05081 & 1.30614 & \color{red}\bf2.26680 & 1.52834 & 1.54028  \\ 
 & total economic costs& 0.67889 & 0.40734 & 0.42732 & \textbf{\textit{0.36805}} & 0.57942 & 0.58338 & 0.42273 & \textbf{0.34457}  \\ 
 & std. total economic costs& 0.03692 & 0.03707 & 0.03661 & 0.03284 & 0.04469 & 0.19144 & 0.04018 & 0.02213  \\ 
 & producers' profits & \color{red}\bf-0.03667 & 0.08520 & 0.11065 & \textbf{\textit{0.14226}} & 0.14174 & \color{red}\bf0.89434 & 0.30625 & \color{blue}{0.40691}  \\ 
 & std. producers' profits & 0.01338 & 0.01984 & 0.01641 & 0.01614 & 0.01953 & 0.31197 & 0.01754 & 0.01836  \\ 
 & grid workload & 0.96603 & \textbf{0.89847} & 0.93219 & \textbf{0.89505} & 1.04123 & 1.42193 & 1.06795 & 1.01569  \\ 
 & std. grid workload & 0.13985 & 0.12508 & 0.12458 & 0.11826 & 0.15021 & 0.33491 & 0.13610 & 0.12944  \\ 
 & curtailment & 0.00000 & 0.02991 & 0.00000 & 0.02823 & 0.00000 & 0.06547 & 0.00000 & 0.03201  \\ 
 & upward balancing & \color{red}\bf0.30591 & 0.08268 & 0.10572 & 0.05605 & 0.20781 & 0.00009 & 0.08169 & 0.00864  \\ 
 & downward balancing & 0.00000 & 0.00004 & 0.00000 & 0.00000 & 0.00000 & 0.23065 & 0.00000 & 0.00889  \\ 
 & $(q_i-\mu_i)/\rho_i\mu_i$ & -0.22589 & -0.10612 & 0.00842 & -0.01501 & -0.22483 & -0.04297 & 0.08519 & 0.07803  \\ 
 & bid price-to-cost ratio $p_i/c_i$ & \textbf{2.73163} & 4.49091 & 6.52608 & \textbf{\textit{4.11396}} & 6.99135 & 10.20656 & 8.69645 & 7.87349
\end{tabular}
\end{center}
\end{footnotesize}
\end{sidewaystable}
\begin{sidewaystable}[p]
\begin{footnotesize}
\begin{center}
\begin{tabular}{ll|rrrr|rrrr}
\color{white}centrality & \multicolumn{1}{l}{}  & \multicolumn{1}{l}{\color{white}don't curtail} & \multicolumn{1}{l}{\color{white}curtail} & \multicolumn{1}{l}{\color{white}don't curtail} & \multicolumn{1}{l}{\color{white}curtail} & \multicolumn{1}{l}{\color{white}don't curtail} & \multicolumn{1}{l}{\color{white}curtail} & \multicolumn{1}{l}{\color{white}don't curtail} & \multicolumn{1}{l}{\color{white}curtail}  \\ 
-0.5 & price & \textbf{0.49363} & 0.96849 & 0.91063 & 0.92646 & 1.23838 & \color{red}\bf2.36502 & 1.45549 & 1.48011  \\ 
 & total economic costs& 0.68711 & 0.39233 & 0.40382 & \textbf{\textit{0.35677}} & 0.55519 & 0.58779 & 0.38721 & \textbf{0.30770}  \\ 
 & std. total economic costs& 0.03894 & 0.04061 & 0.04014 & 0.03495 & 0.04759 & 0.16097 & 0.04466 & 0.02421  \\ 
 & producers' profits & \color{red}\bf-0.04955 & 0.04783 & 0.07391 & \textbf{\textit{0.10305}} & 0.14026 & \color{red}\bf1.00328 & 0.29239 & \color{blue}{0.40000}  \\ 
 & std. producers' profits & 0.01023 & 0.02839 & 0.01806 & 0.01900 & 0.02243 & 0.29001 & 0.01794 & 0.02319  \\ 
 & grid workload & 0.98527 & \textbf{\textit{0.80879}} & 0.81490 & \textbf{0.78057} & 0.96520 & 1.28332 & 0.91835 & 0.83154  \\ 
 & std. grid workload & 0.12699 & 0.10531 & 0.09217 & 0.09410 & 0.13563 & 0.26973 & 0.14056 & 0.11437  \\ 
 & curtailment & 0.00000 & 0.03622 & 0.00000 & 0.03593 & 0.00000 & 0.09426 & 0.00000 & 0.04276  \\ 
 & upward balancing & \color{red}\bf0.32013 & 0.09993 & 0.11872 & 0.07830 & 0.20966 & 0.00032 & 0.08401 & 0.01489  \\ 
 & downward balancing & 0.00000 & 0.00001 & 0.00000 & 0.00000 & 0.00000 & 0.28077 & 0.00002 & 0.00583  \\ 
 & $(q_i-\mu_i)/\rho_i\mu_i$ & -0.20856 & -0.09722 & 0.00506 & -0.01985 & -0.20794 & -0.02952 & 0.08030 & 0.07368  \\ 
 & bid price-to-cost ratio $p_i/c_i$ & \textbf{3.18489} & 5.13785 & 8.61435 & \textbf{\textit{4.70348}} & 8.79295 & 13.49640 & 11.41692 & 10.00518 \\ \midrule
-1 & price & \textbf{0.44077} & 0.79917 & 0.77089 & \textbf{\textit{0.75874}} & 1.13505 & \color{red}\bf2.05625 & 1.31326 & 1.32840 \\ 
 & total economic costs& 0.63303 & 0.35337 & 0.35574 & \textbf{\textit{0.32651}} & 0.49881 & 0.46149 & 0.36595 & \textbf{0.28619}  \\ 
 & std. total economic costs& 0.15669 & 0.06126 & 0.07288 & 0.05162 & 0.09085 & 0.14501 & 0.05254 & 0.03019  \\ 
 & producers' profits & \color{red}\bf-0.04854 & 0.00925 & 0.04089 & \textbf{\textit{0.05118}} & 0.14386 & \color{red}\bf0.78859 & 0.25067 & \color{blue}{0.33845}  \\ 
 & std. producers' profits & 0.04715 & 0.04348 & 0.01912 & 0.02765 & 0.06272 & 0.27698 & 0.03040 & 0.02223  \\ 
 & grid workload & 0.95493 & 0.80361 & 0.86136 & \textbf{0.78869} & 0.95681 & 1.12602 & 0.90801 & \textbf{\textit{0.80521}}  \\ 
 & std. grid workload & 0.16581 & 0.10706 & 0.09843 & 0.11226 & 0.15517 & 0.22130 & 0.16142 & 0.14145  \\ 
 & curtailment & 0.00000 & 0.04751 & 0.00000 & 0.04487 & 0.00000 & 0.11157 & 0.00000 & 0.04792  \\ 
 & upward balancing & \color{red}\bf0.29178 & 0.09387 & 0.10582 & 0.08024 & 0.18455 & 0.00115 & 0.09190 & 0.02957  \\ 
 & downward balancing & 0.00081 & 0.00030 & 0.00055 & 0.00053 & 0.00044 & 0.21056 & 0.00052 & 0.00074  \\ 
 & $(q_i-\mu_i)/\rho_i\mu_i$ & -0.23269 & -0.11270 & 0.00912 & -0.03703 & -0.23992 & -0.02526 & 0.10508 & 0.08204  \\ 
 & bid price-to-cost ratio $p_i/c_i$ & \textbf{3.51436} & 4.74751 & 8.34417 & \textbf{\textit{4.61745}} & 9.67136 & 14.62034 & 12.16332 & 10.87470  \\ \midrule
all & price   & \textbf{0.62495} & 1.06822 & 1.01734 & 1.02063 & 1.29958 & \color{red}\bf2.17382 & 1.50435 & 1.52118  \\ 
 & total economic costs  & 0.66027 & 0.41095 & 0.41669 & \textbf{\textit{0.36963}} & 0.56793 & 0.52624 & 0.41019 & \textbf{0.34287}  \\ 
 & std. total economic costs  & 0.06926 & 0.05277 & 0.04879 & 0.04159 & 0.05739 & 0.16733 & 0.04887 & 0.04524  \\ 
 & producers' profits   & \color{red}\bf-0.03309 & 0.07900 & 0.10412 & \textbf{\textit{0.13224}} & 0.14631 & \color{red}\bf0.81035 & 0.30225 & \color{blue}{0.40211}  \\ 
 & std. producers' profits   & 0.02646 & 0.05275 & 0.04128 & 0.04906 & 0.03016 & 0.29618 & 0.03112 & 0.03226  \\ 
 & grid workload   & 0.96249 & \textbf{0.90007} & 0.95591 & \textbf{0.90254} & 1.05353 & 1.30934 & 1.03098 & 0.98037  \\ 
 & std. grid workload   & 0.15382 & 0.15631 & 0.15673 & 0.15551 & 0.17851 & 0.33428 & 0.17115 & 0.18755  \\ 
 & curtailment  & 0.00000 & 0.03041 & 0.00000 & 0.03061 & 0.00000 & 0.06819 & 0.00000 & 0.03483  \\ 
 & upward balancing  & \color{red}\bf0.29114 & 0.09003 & 0.10216 & 0.06062 & 0.20148 & 0.00073 & 0.07773 & 0.01079  \\ 
 & downward balancing & 0.00009 & 0.00011 & 0.00006 & 0.00015 & 0.00005 & 0.18835 & 0.00008 & 0.01076  \\ 
 & $(q_i-\mu_i)/\rho_i\mu_i$  & -0.23453 & -0.10850 & 0.00377 & -0.02055 & -0.23273 & -0.04645 & 0.07848 & 0.07033  \\ 
 & bid price-to-cost ratio $p_i/c_i$ & \textbf{2.73045} & 4.14358 & 6.20152 & \textbf{\textit{3.79583}} & 6.78798 & 9.93891 & 8.39833 & 7.60097  \\ \bottomrule
\end{tabular}
\end{center}
\caption{\label{tbl:results}
Aggregate simulation results by degree of centralization and market design variant.
Averages and some standard deviations, 
each based on a sample of size $500$ (10 networks with 50 time steps).
\textbf{Boldface} and \textbf{\color{red} red boldface} mark best and extreme values in a row,
\textbf{\em boldface italics} 2nd best values or values considered to be a good compromise, {\color{blue} blue} values considered too large.
}
\end{footnotesize}
\end{sidewaystable}

\def\hgt{0.18\textheight}
\begin{figure}[p]
\begin{center}
\begin{tabular}{cc}
	\includegraphics[height=\hgt]{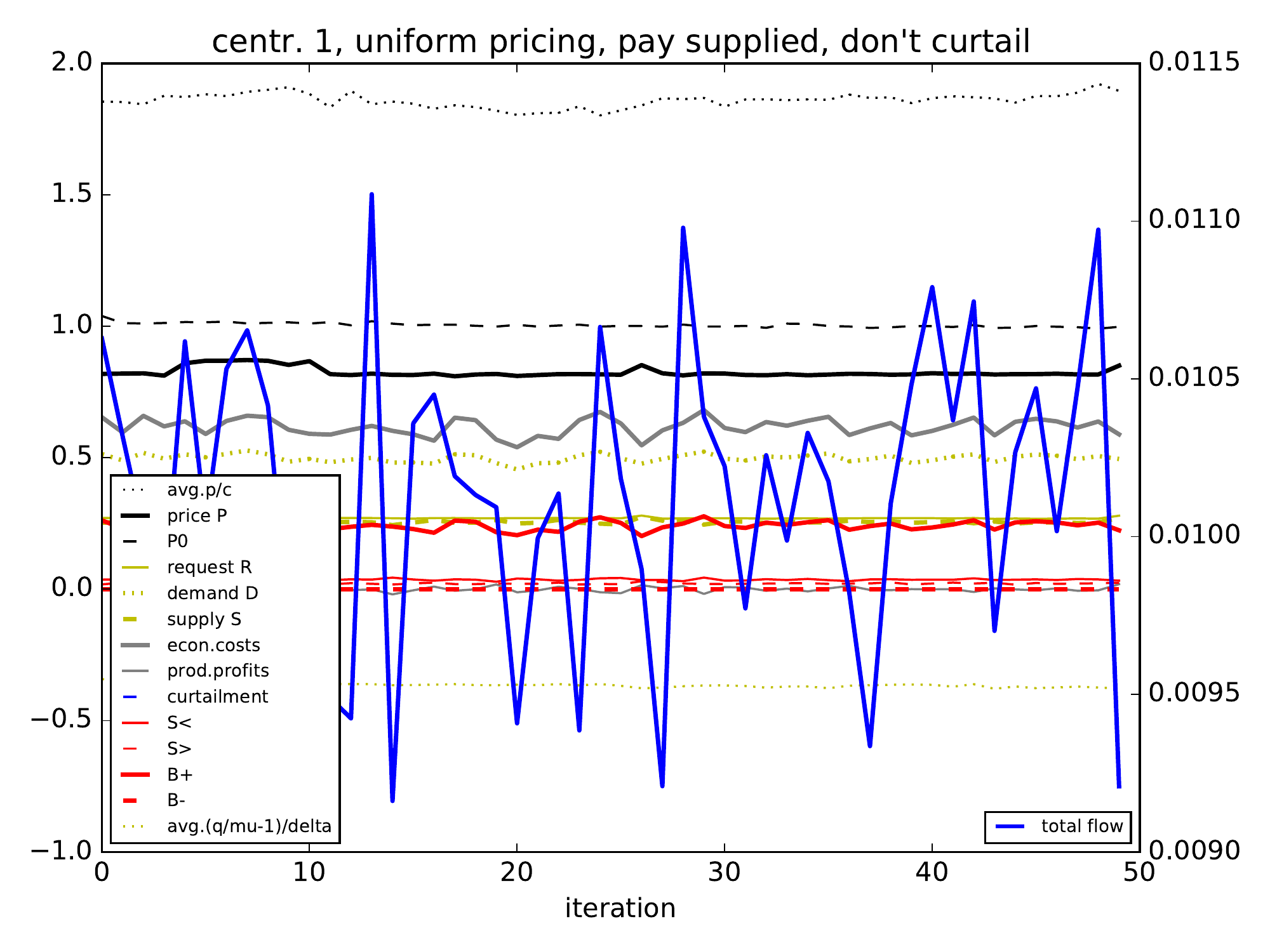} &
	\includegraphics[height=\hgt]{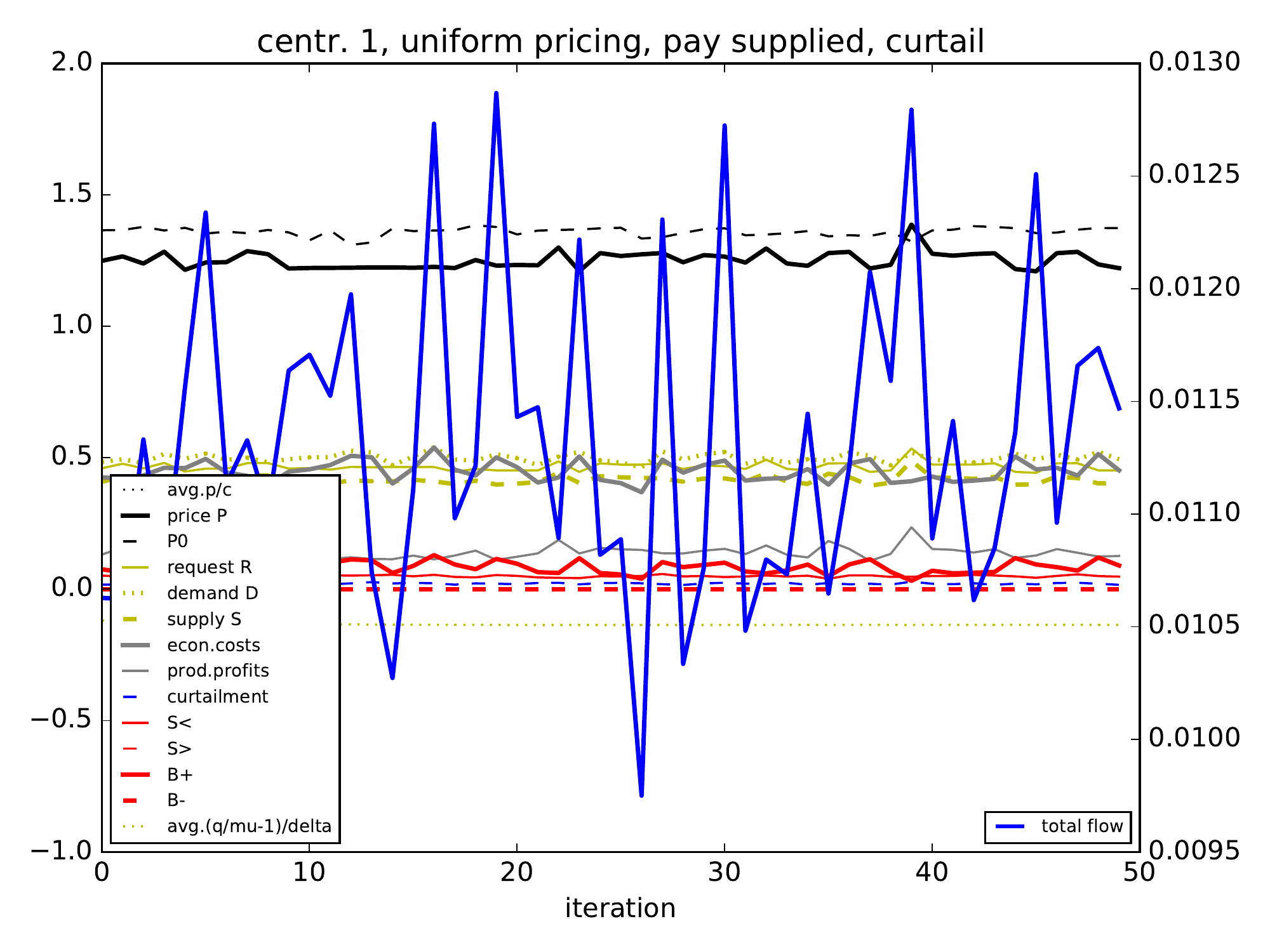} \\
	\includegraphics[height=\hgt]{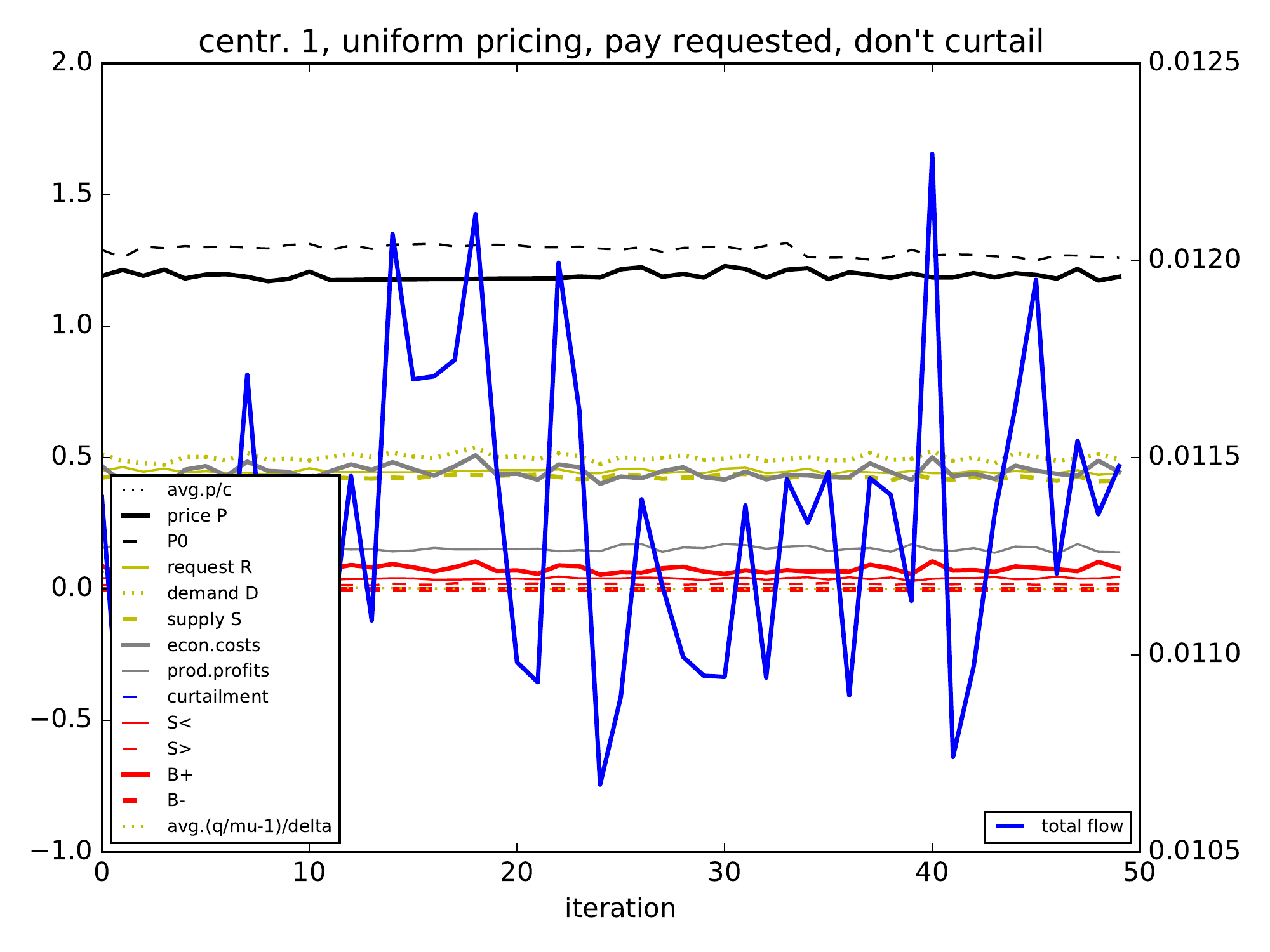} &
	\includegraphics[height=\hgt]{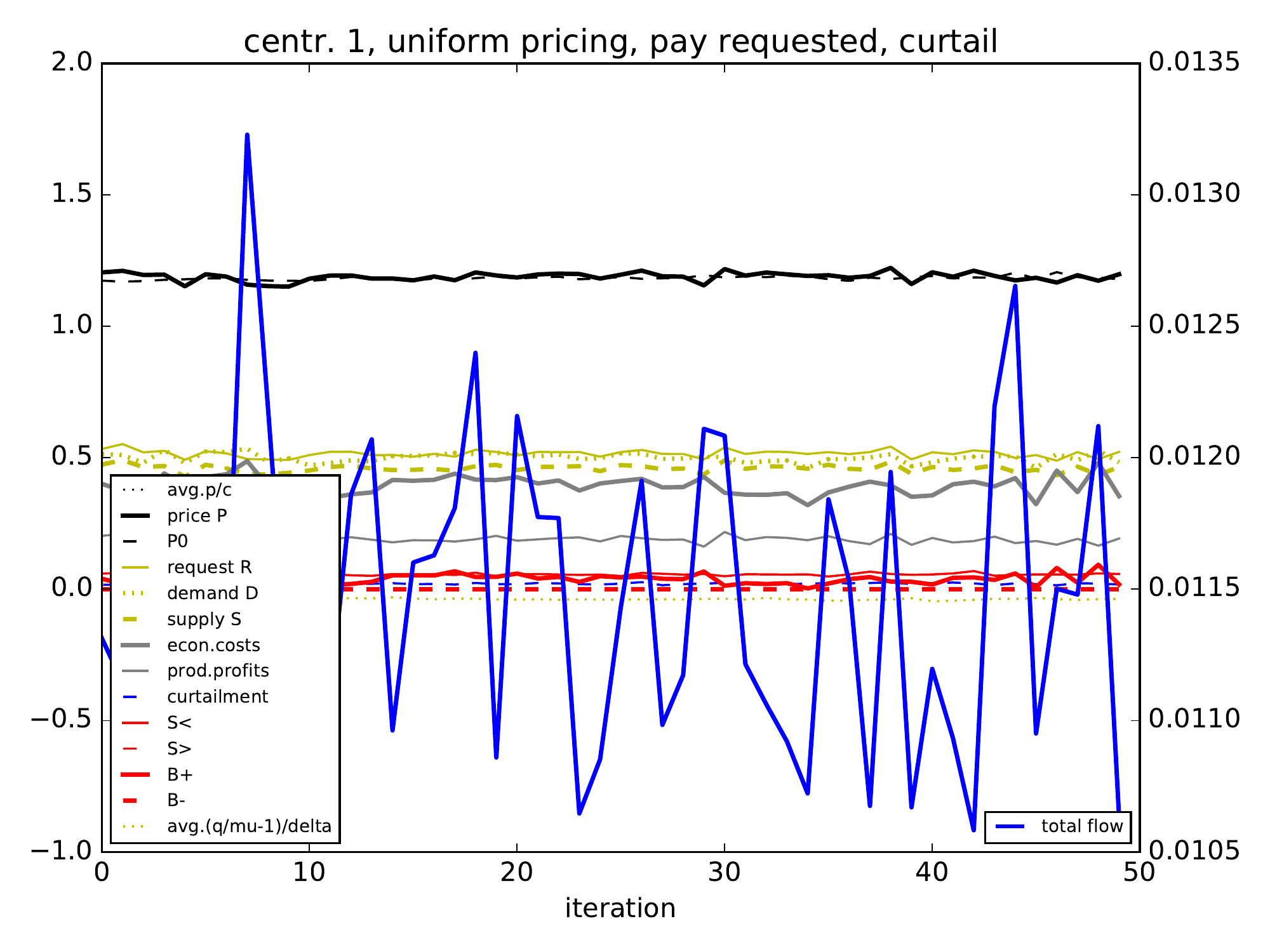} \\
	\includegraphics[height=\hgt]{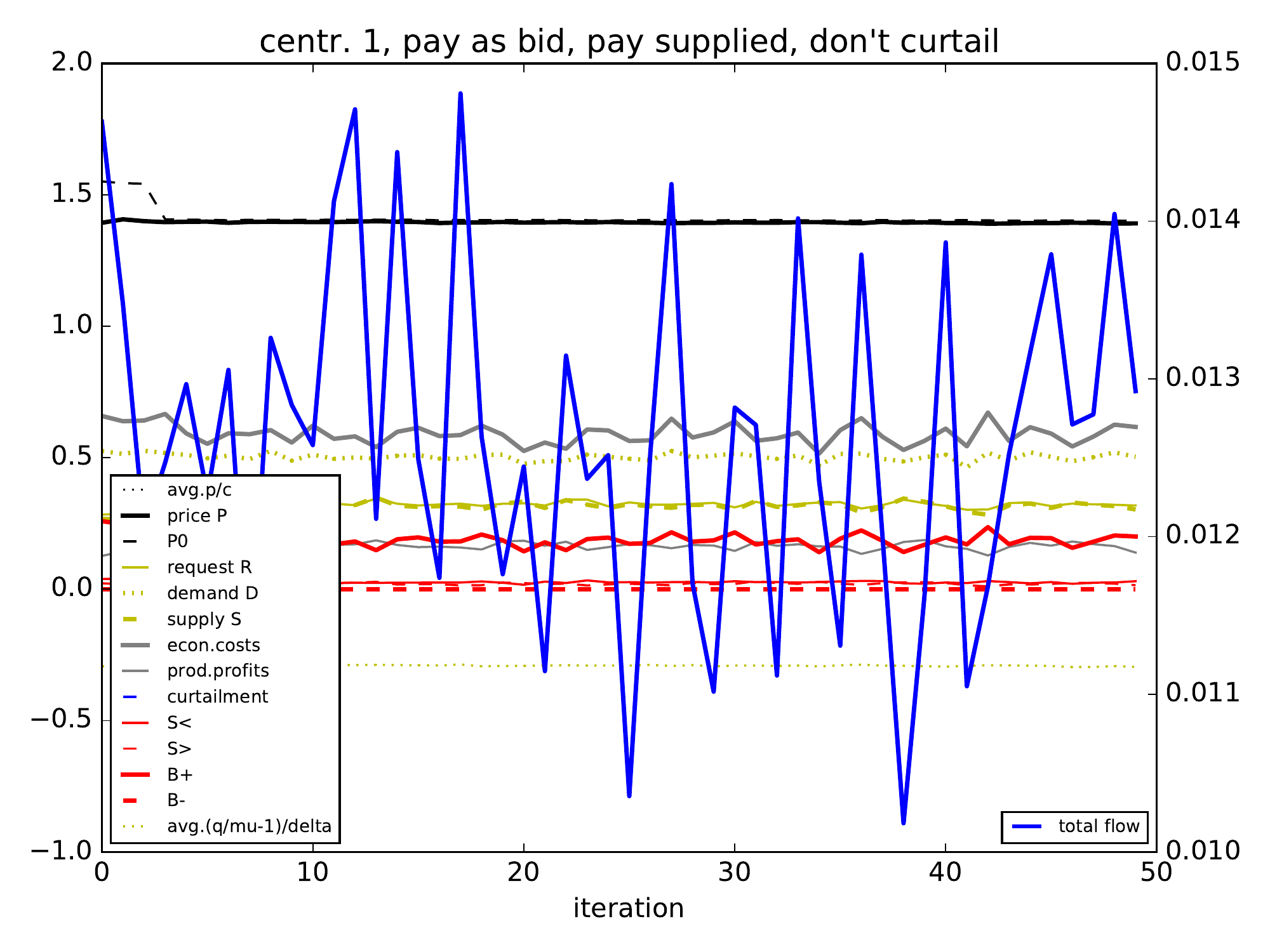} &
	\includegraphics[height=\hgt]{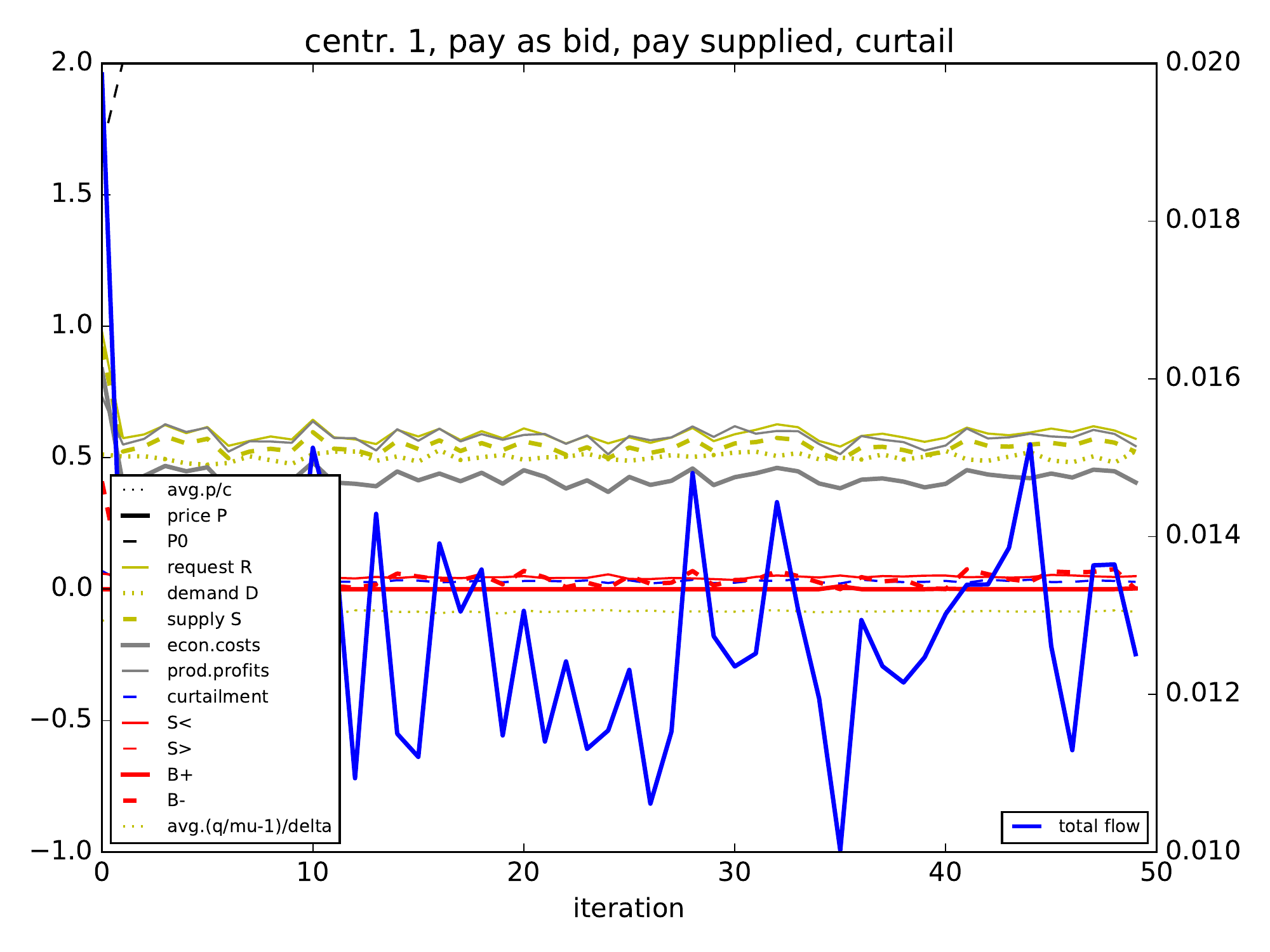} \\
	\includegraphics[height=\hgt]{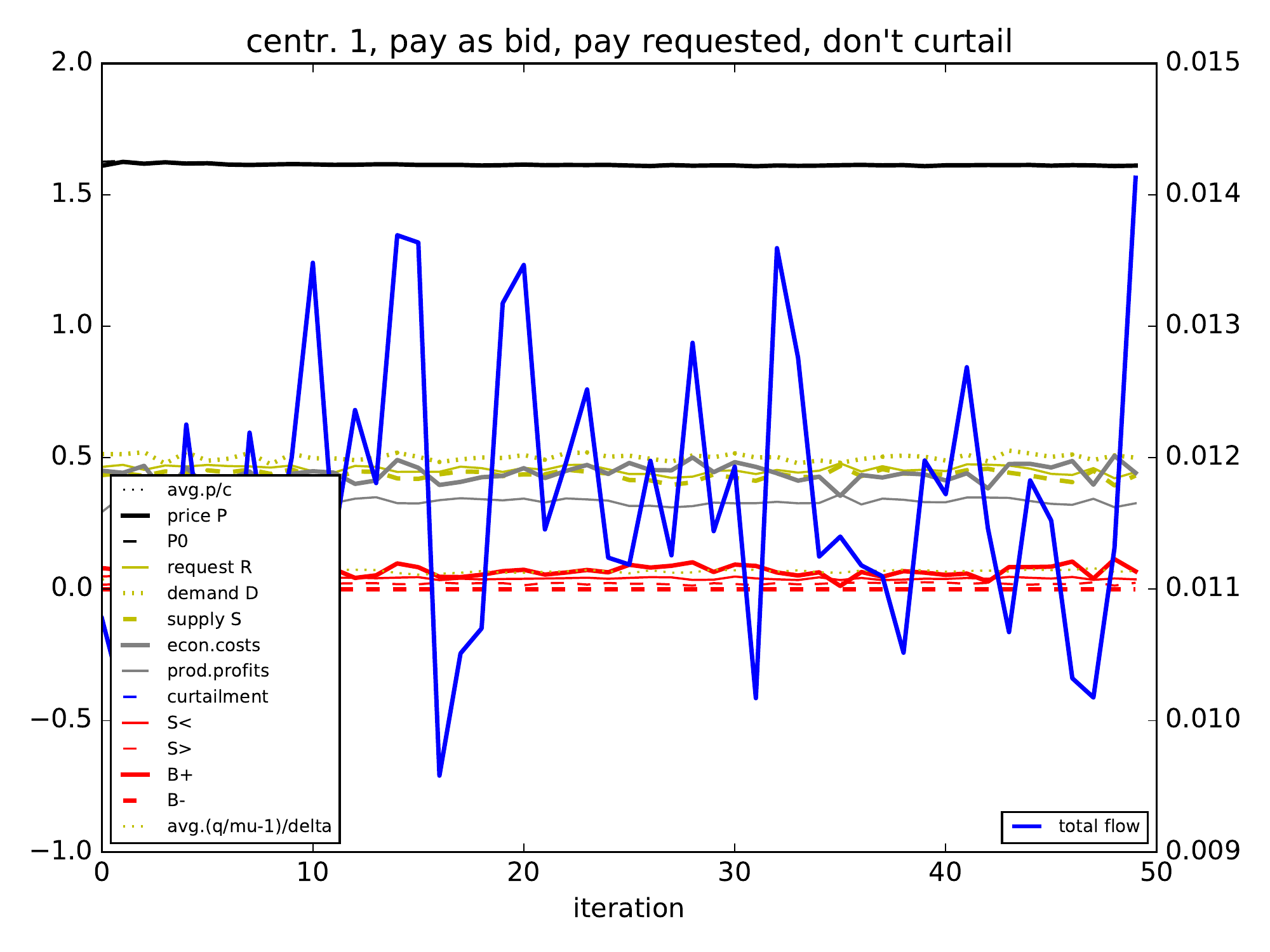} &
	\includegraphics[height=\hgt]{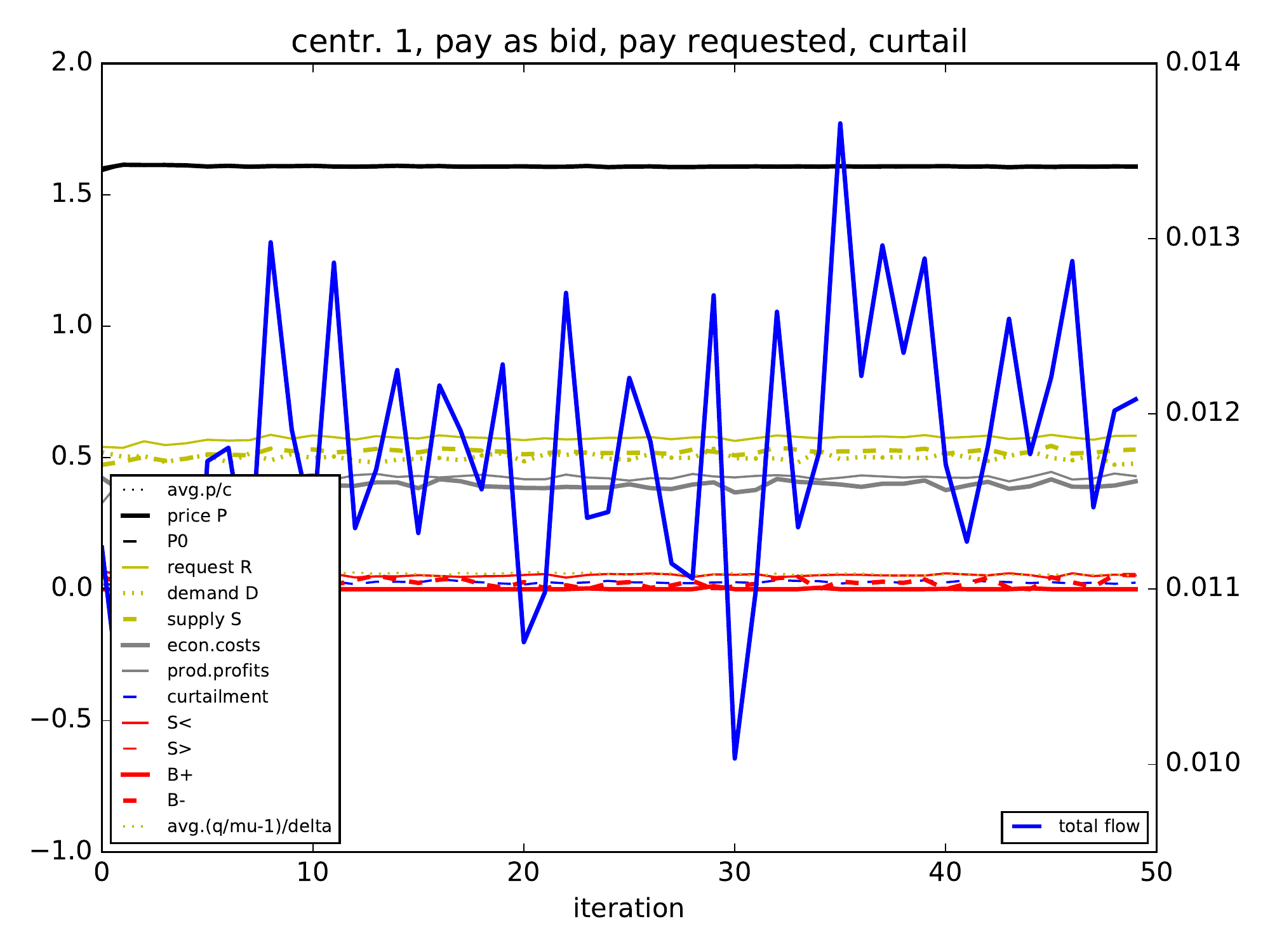} \\
\end{tabular}
\end{center}
\caption{\label{fig:trajectories1}
Sample trajectories for all eight market design variants for the centralized casee $Z=1$.
In all cases, bidding and price-setting behaviour converges almost immediately to a strategic equilibrium. 
}
\end{figure}

\def\hgt{0.18\textheight}
\begin{figure}[p]
\begin{center}
\begin{tabular}{cc}
	\includegraphics[height=\hgt]{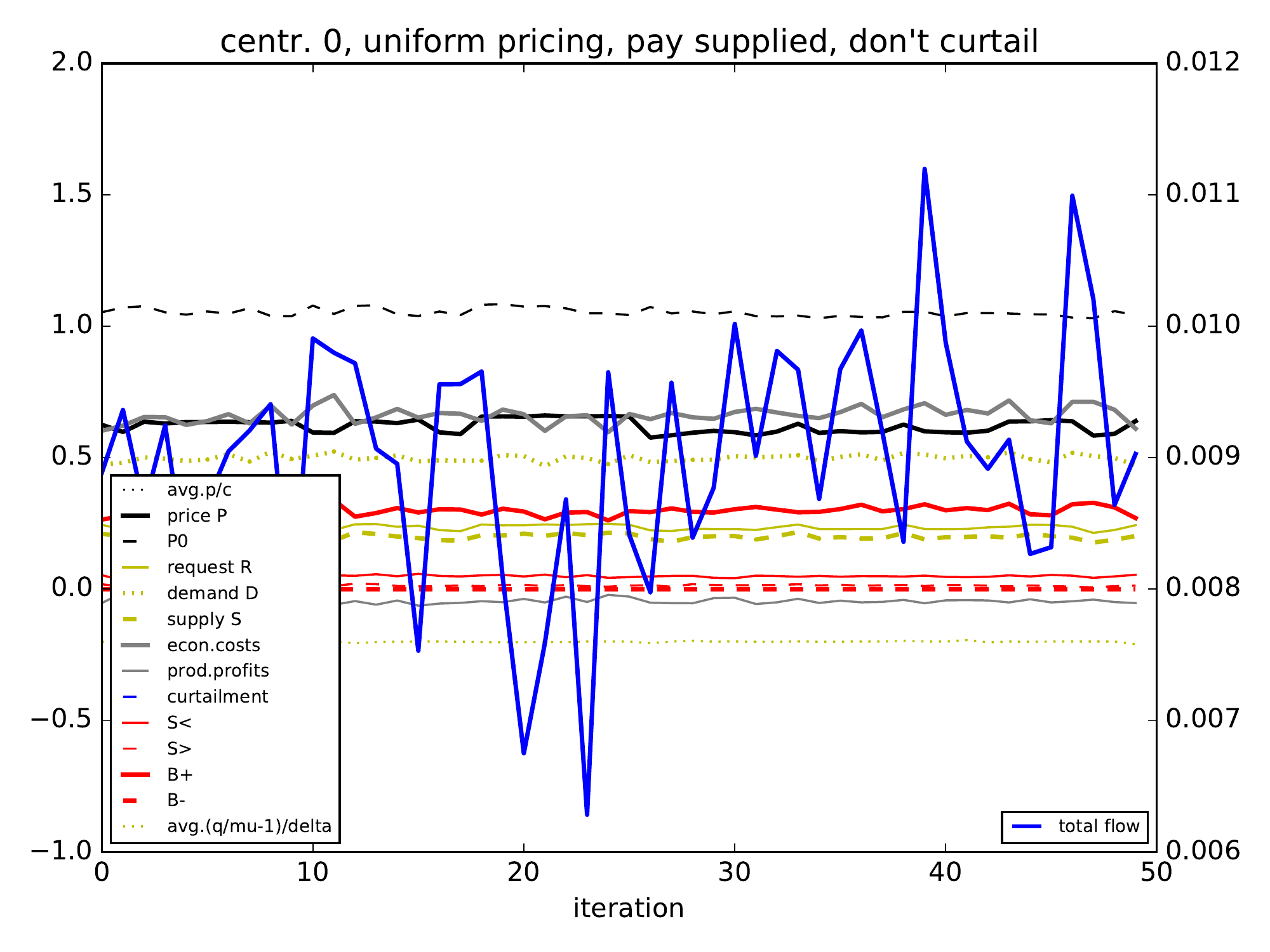} &
	\includegraphics[height=\hgt]{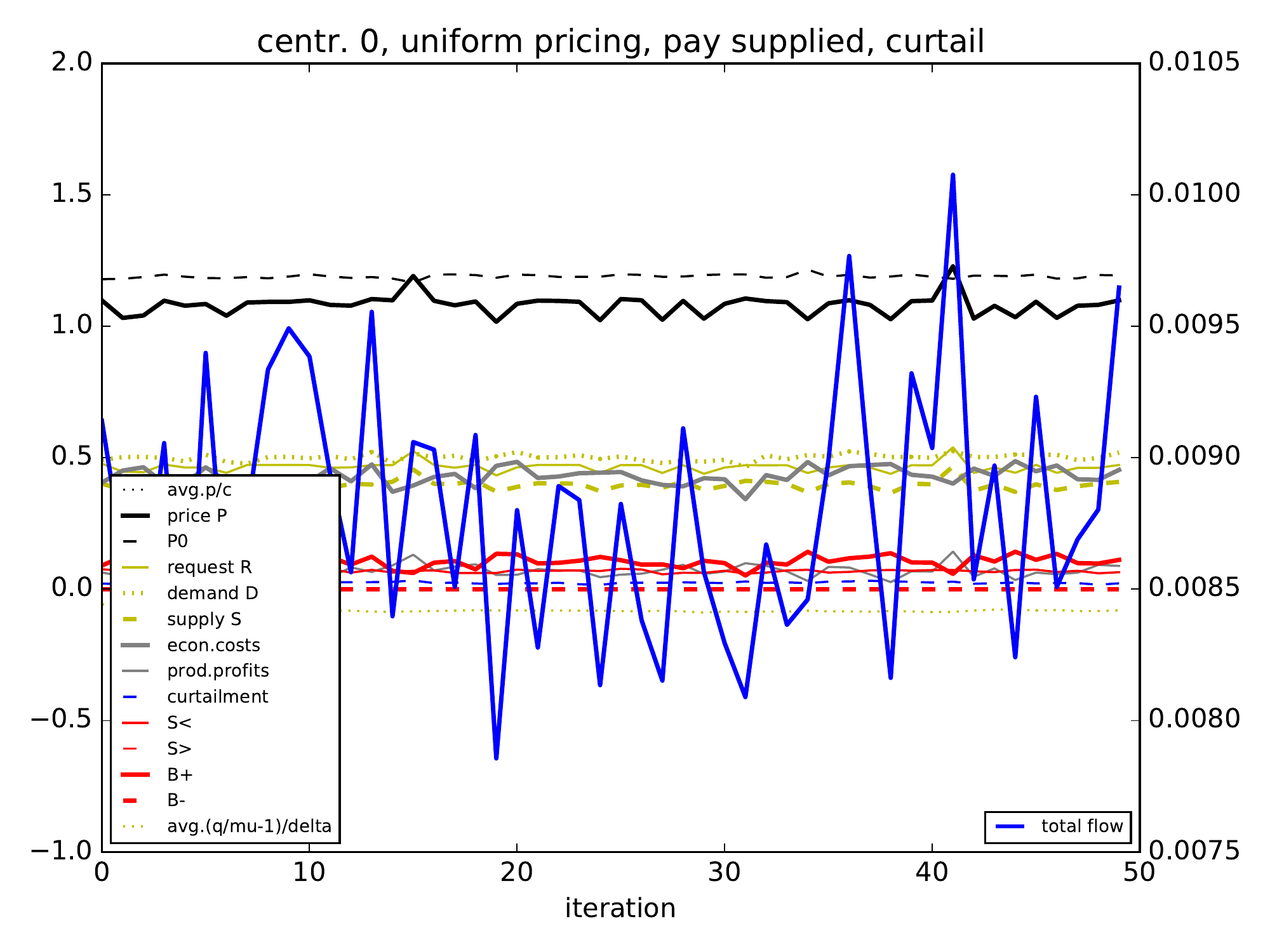} \\
	\includegraphics[height=\hgt]{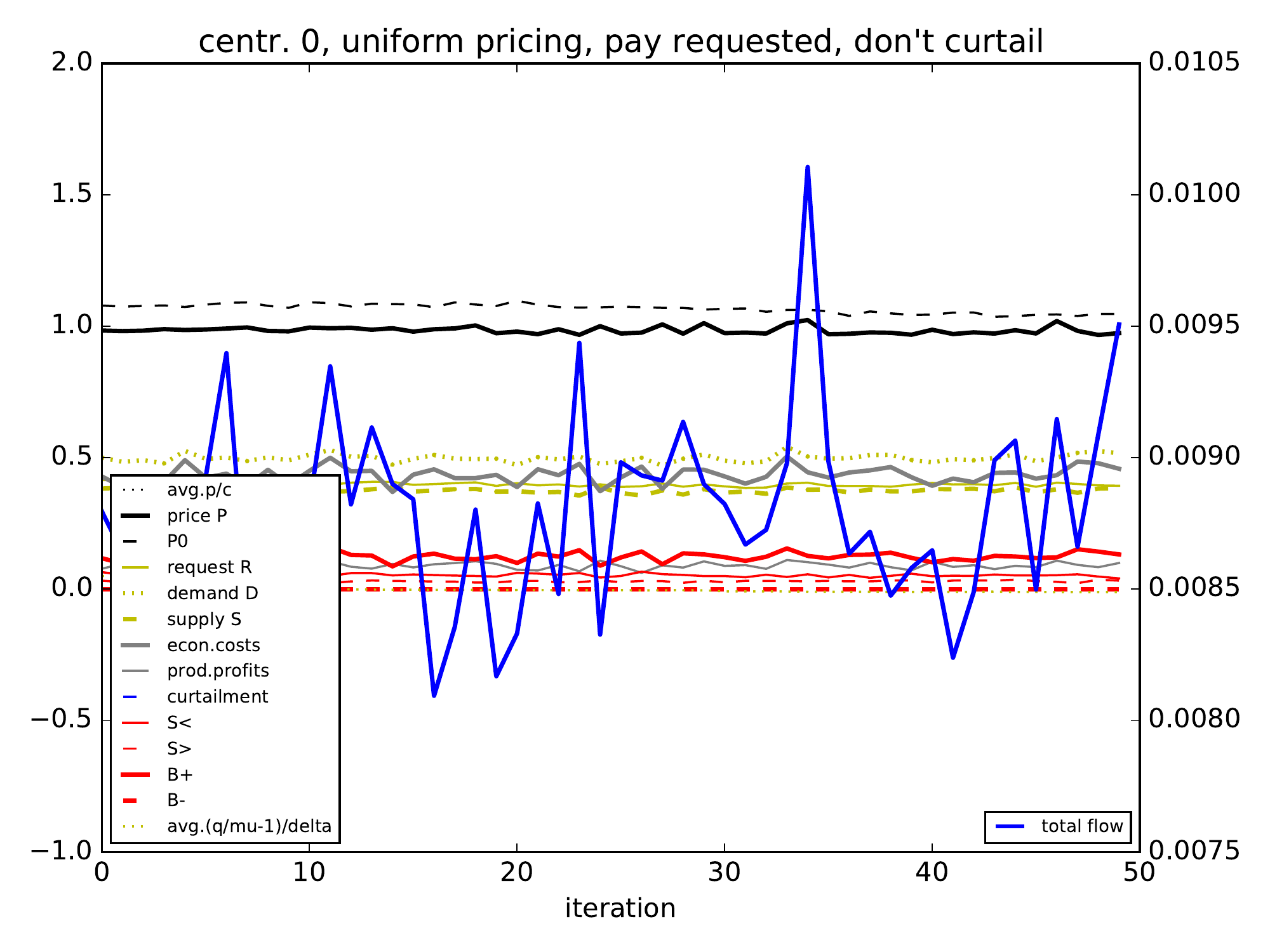} &
	\includegraphics[height=\hgt]{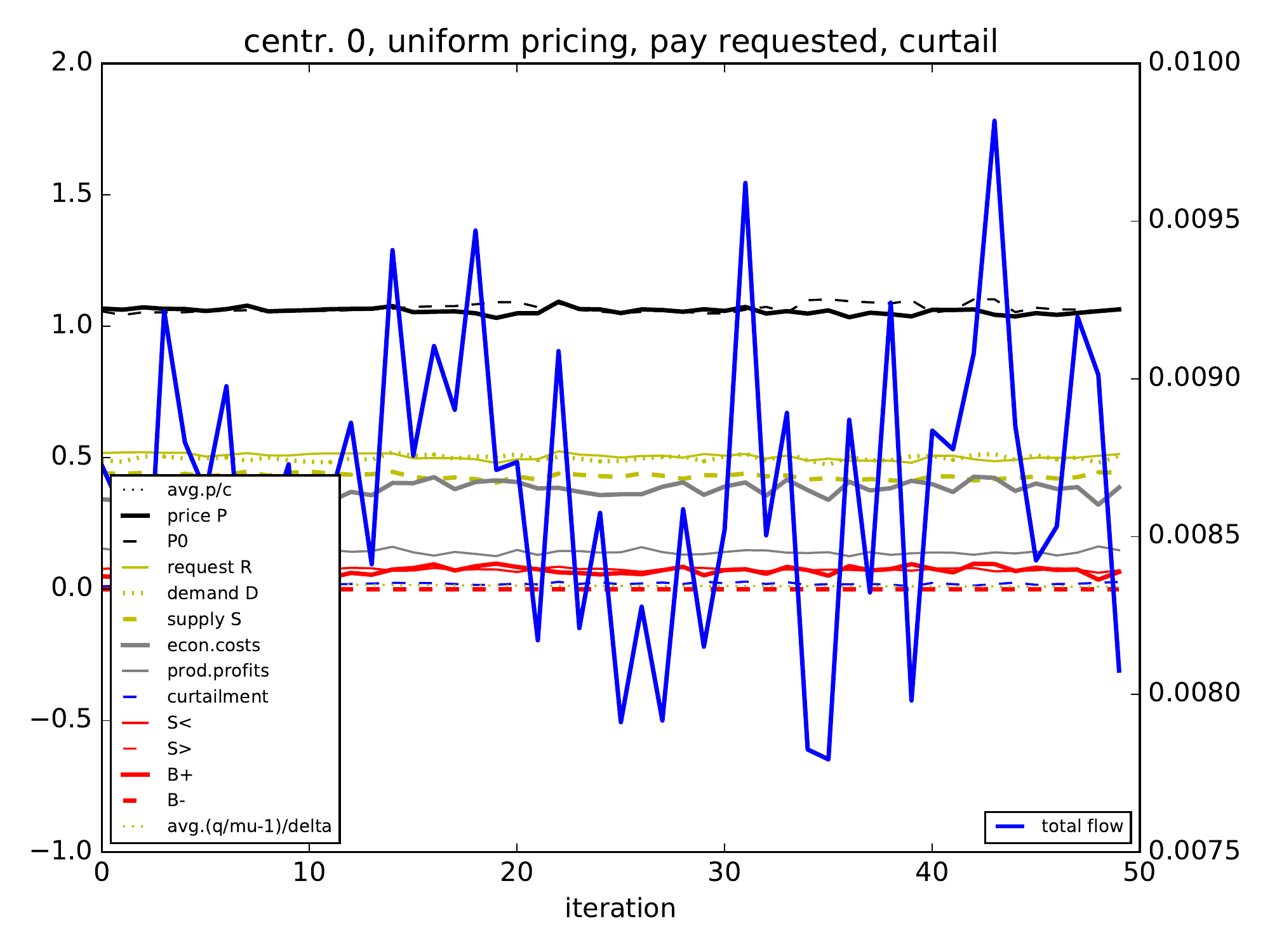} \\
	\multirow{2}{*}[1.8cm]{\includegraphics[height=\hgt]{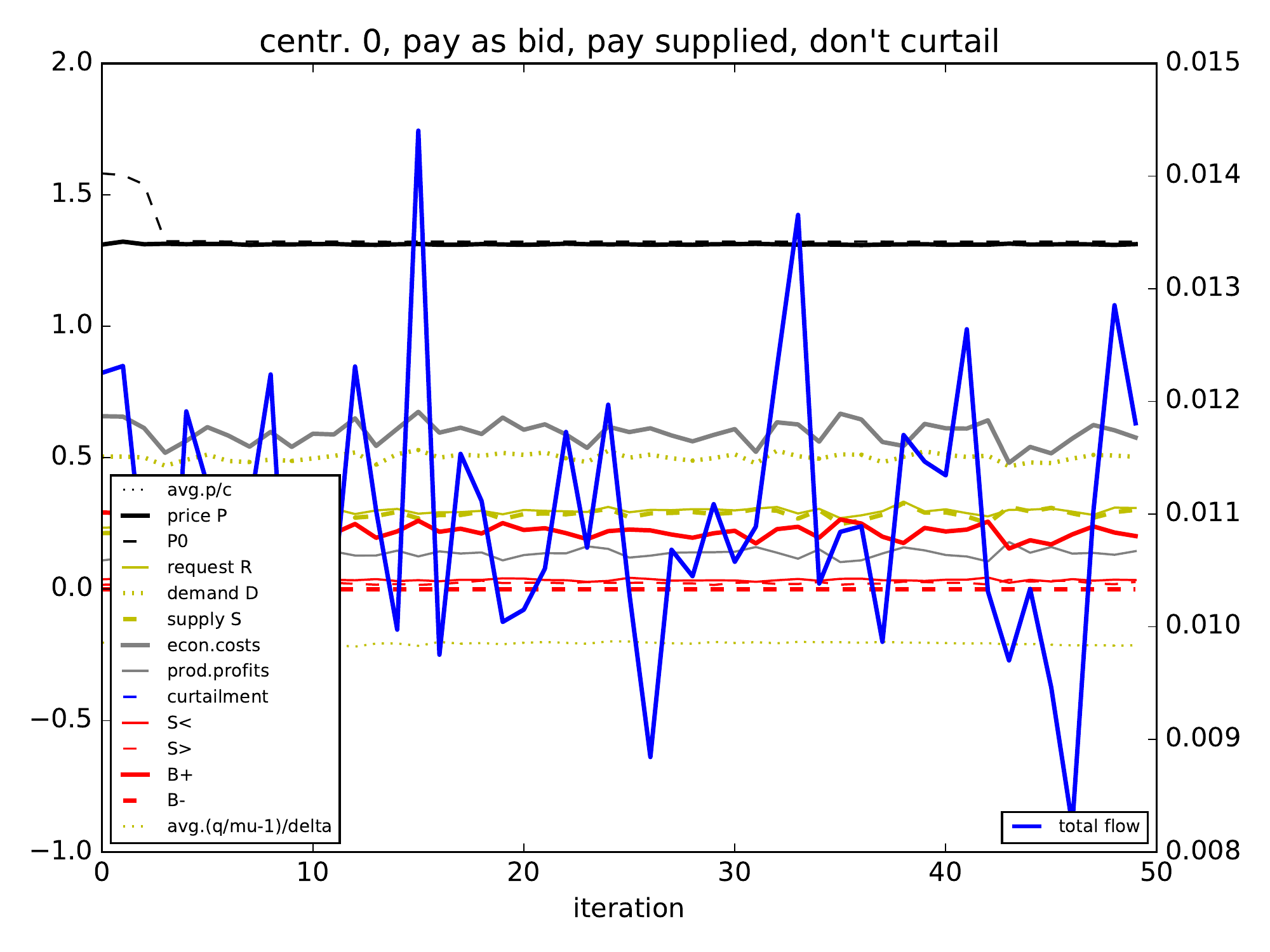}} &
	\includegraphics[height=\hgt]{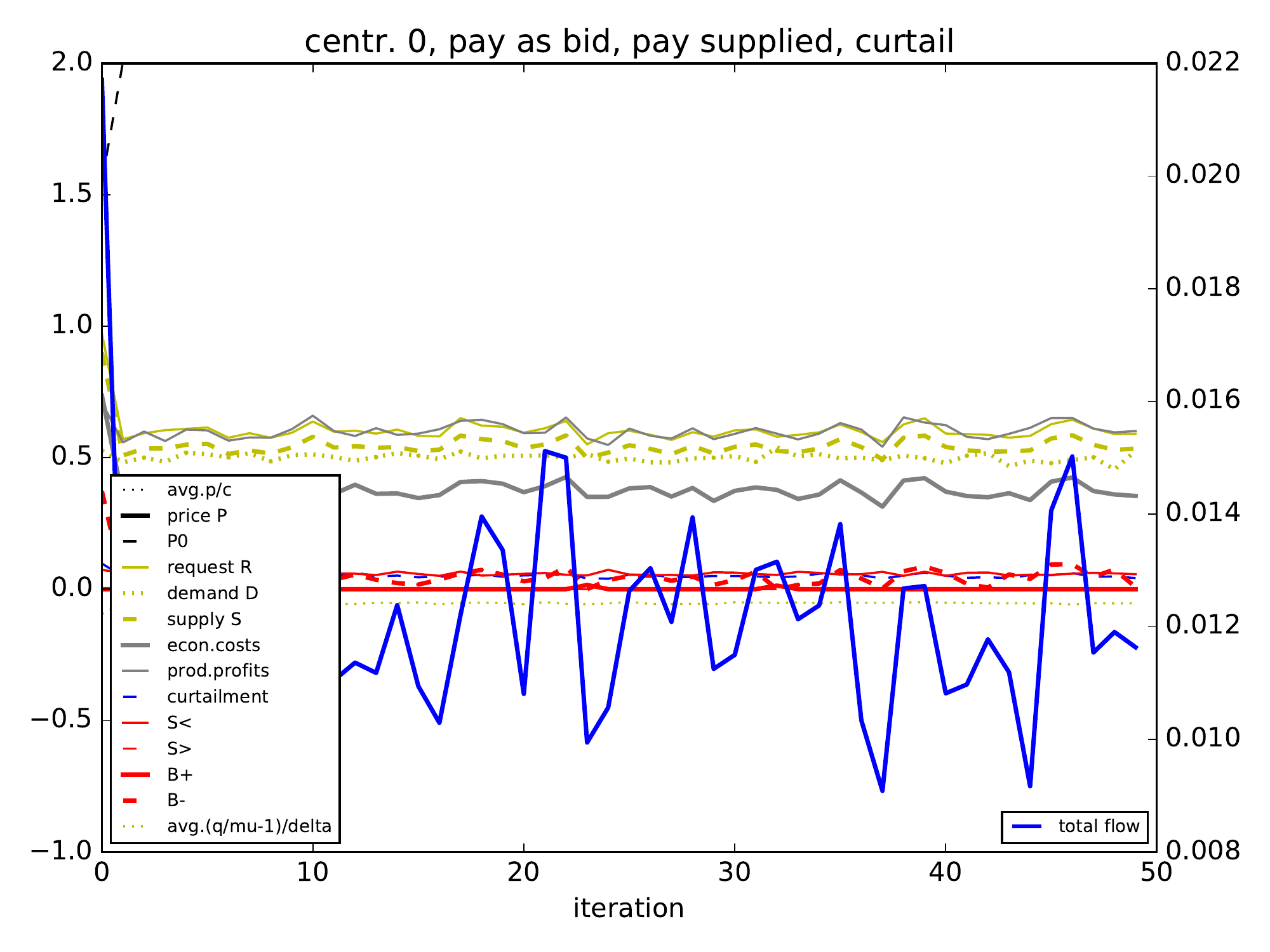} \\
	&
	\includegraphics[height=\hgt]{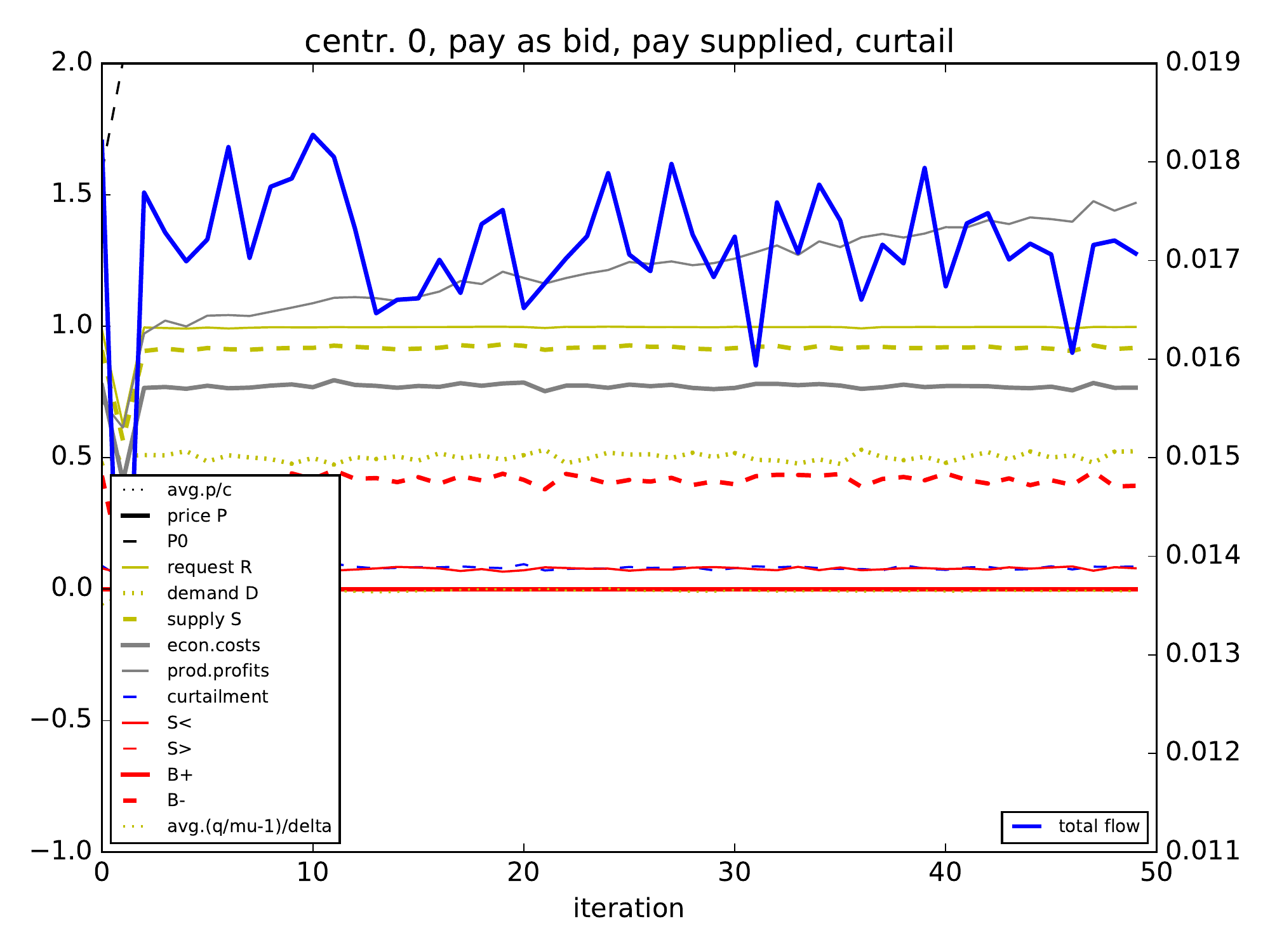} \\
	\includegraphics[height=\hgt]{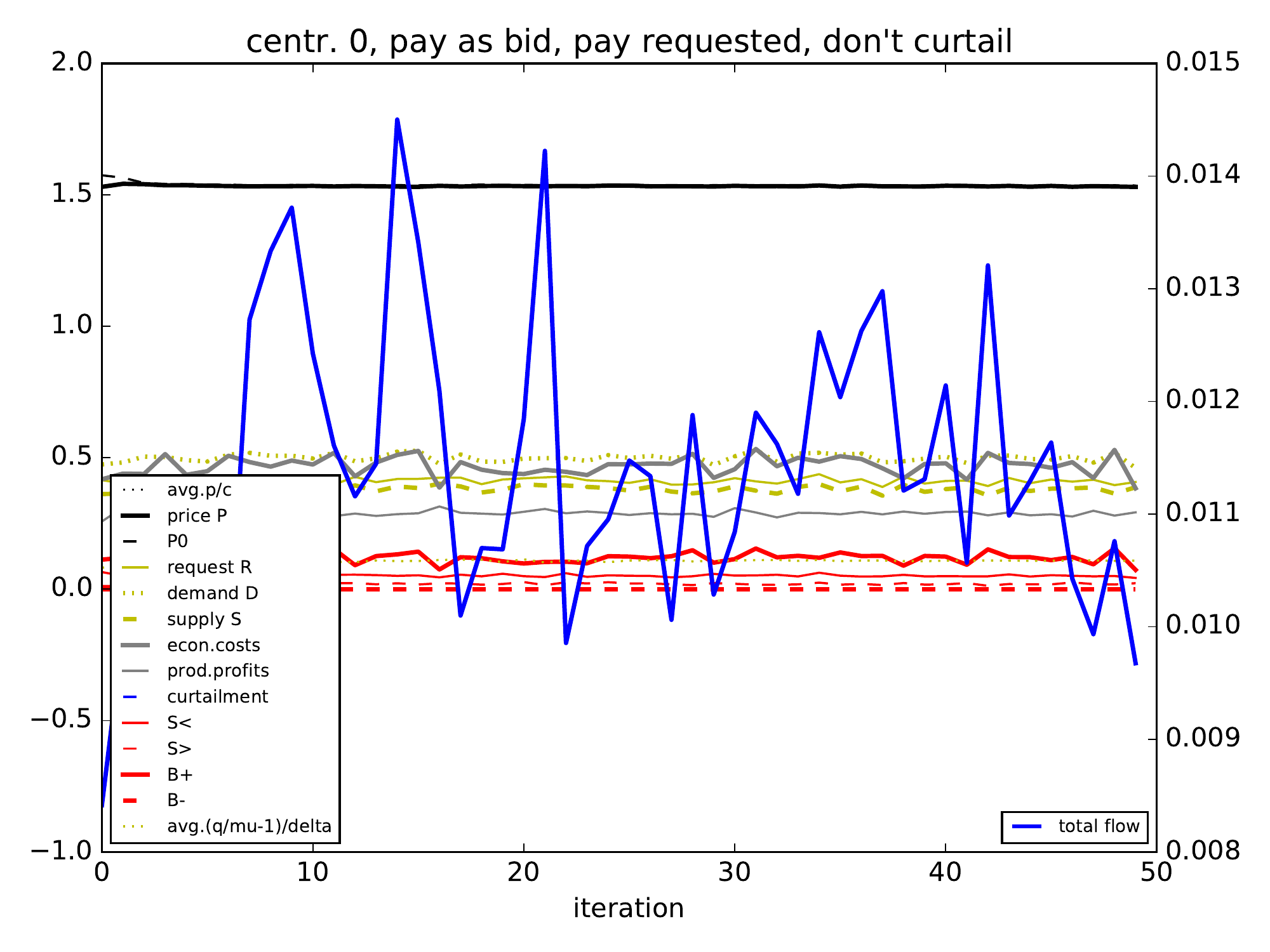} &
	\includegraphics[height=\hgt]{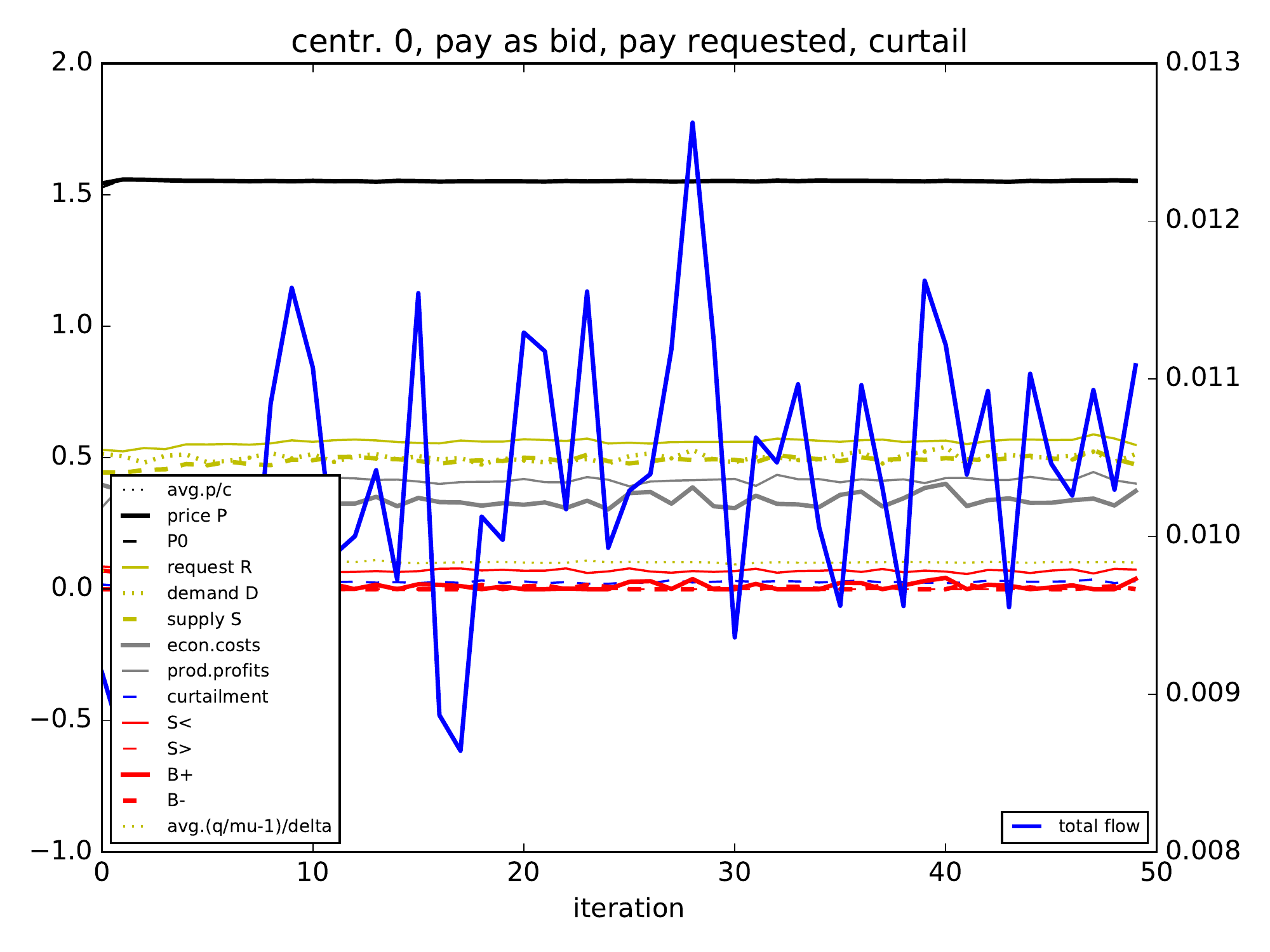} \\
\end{tabular}
\end{center}
\caption{\label{fig:trajectories0}
Sample trajectories for all eight market design variants for no centralization or decentralization ($Z=0$).
Now, in the ``pay as bid, pay supplied, curtail'' variant, 
an additional run with non-converging but increasing price and profits is shown.
In all other cases, bidding and price-setting behaviour converges almost immediately to a strategic equilibrium. 
}
\end{figure}

\def\hgt{0.18\textheight}
\begin{figure}[p]
\begin{center}
\begin{tabular}{cc}
	\includegraphics[height=\hgt]{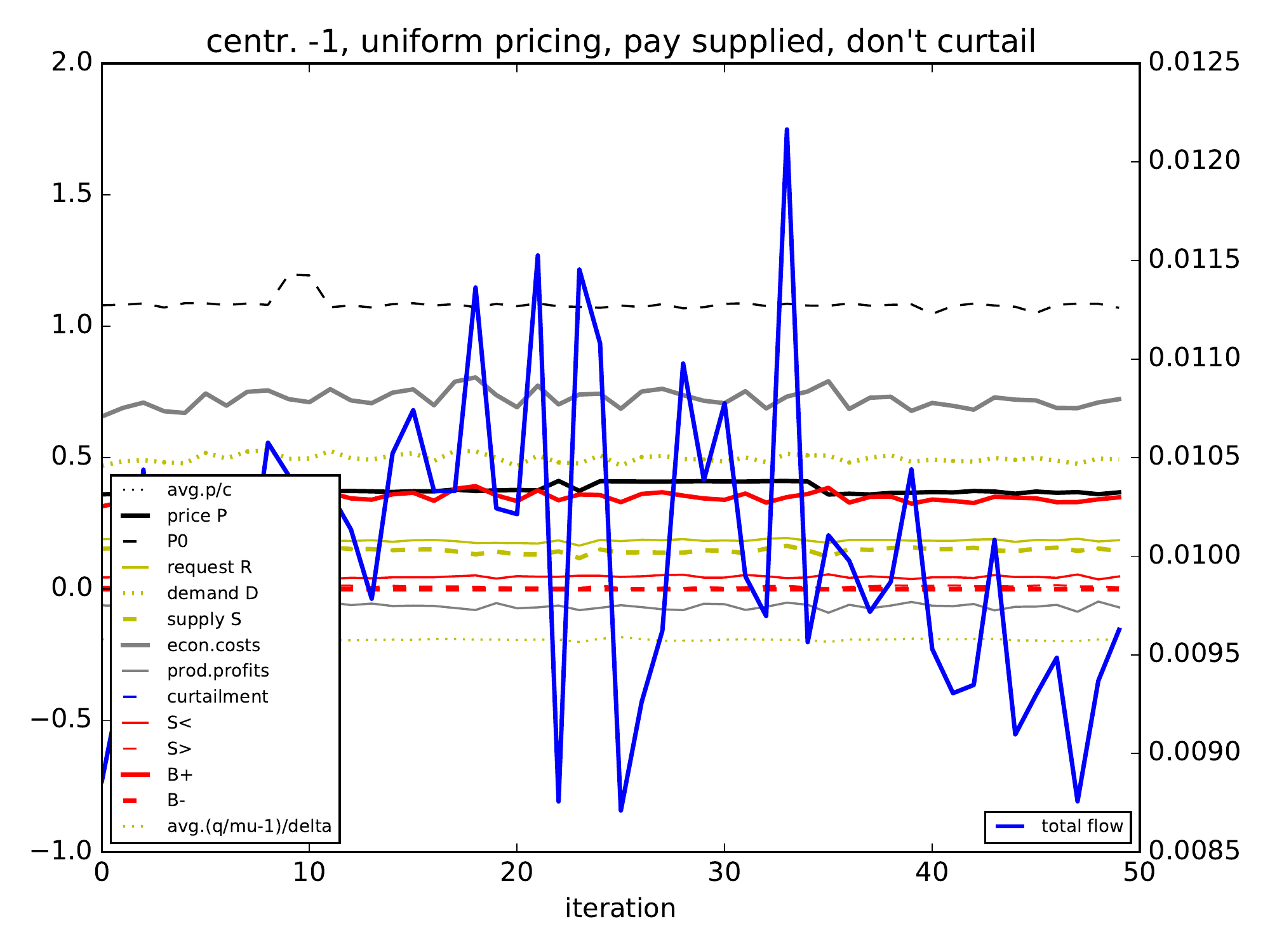} &
	\includegraphics[height=\hgt]{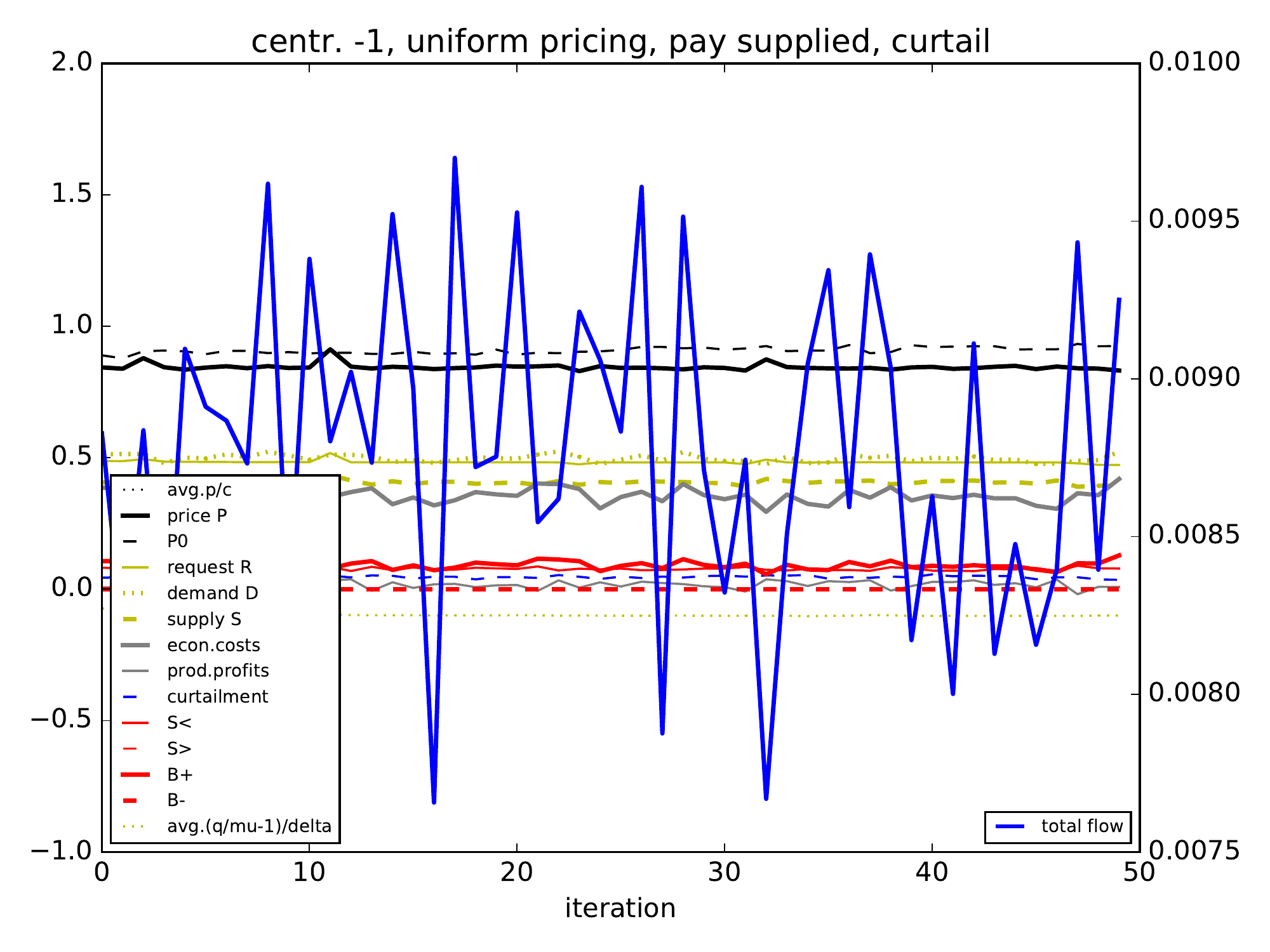} \\
	\includegraphics[height=\hgt]{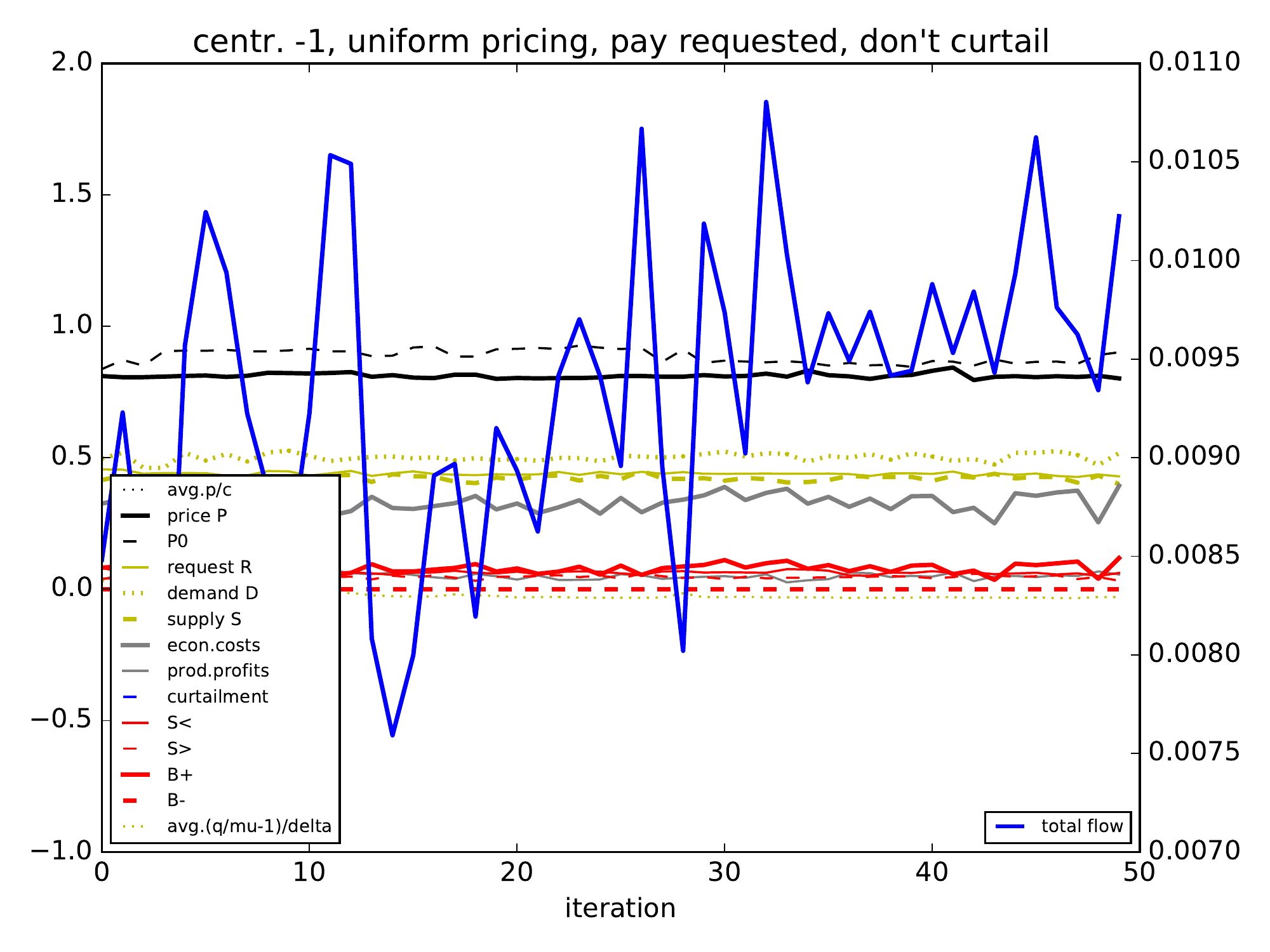} &
	\includegraphics[height=\hgt]{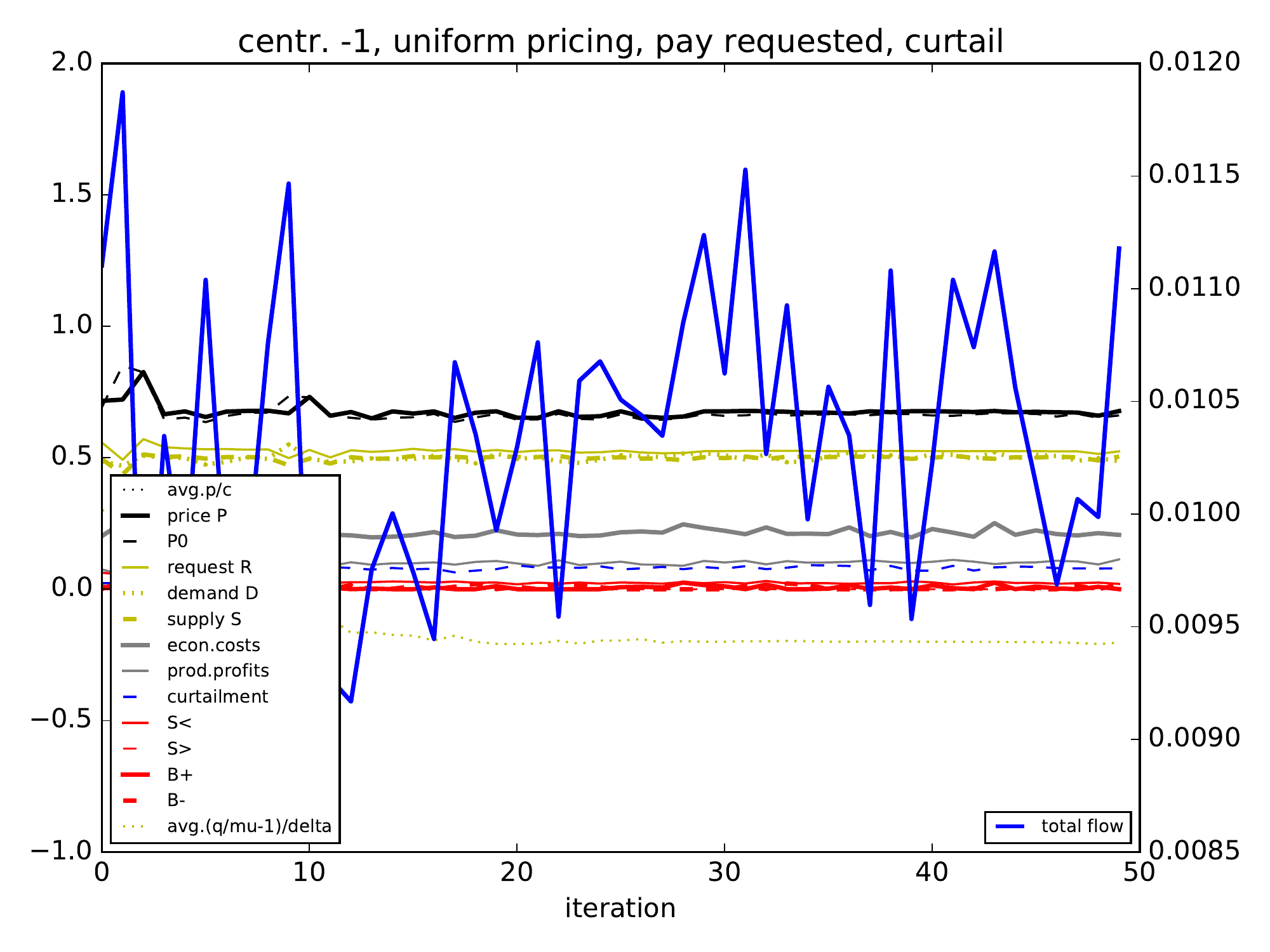} \\
	\multirow{2}{*}[1.8cm]{\includegraphics[height=\hgt]{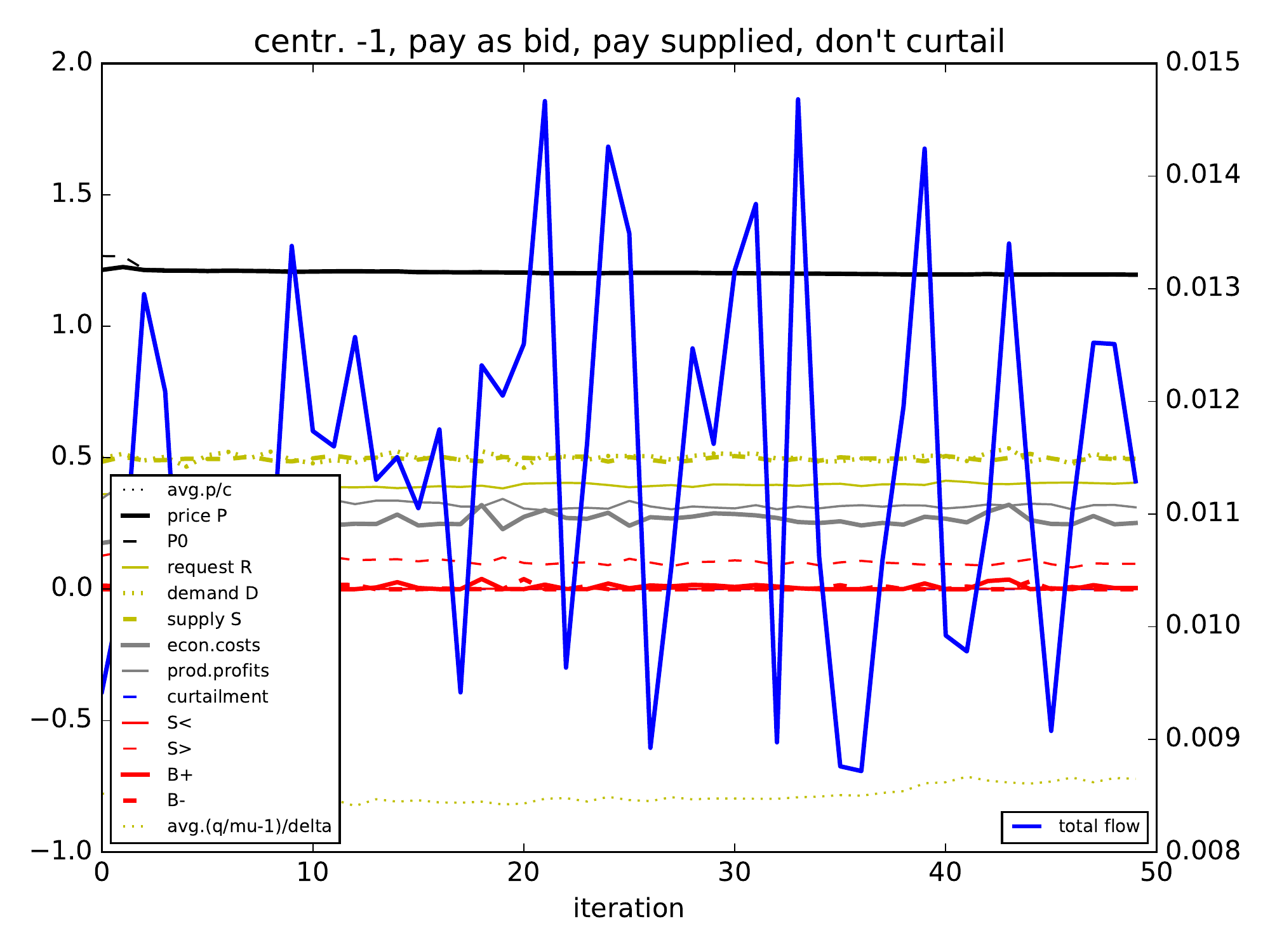}} &
	\includegraphics[height=\hgt]{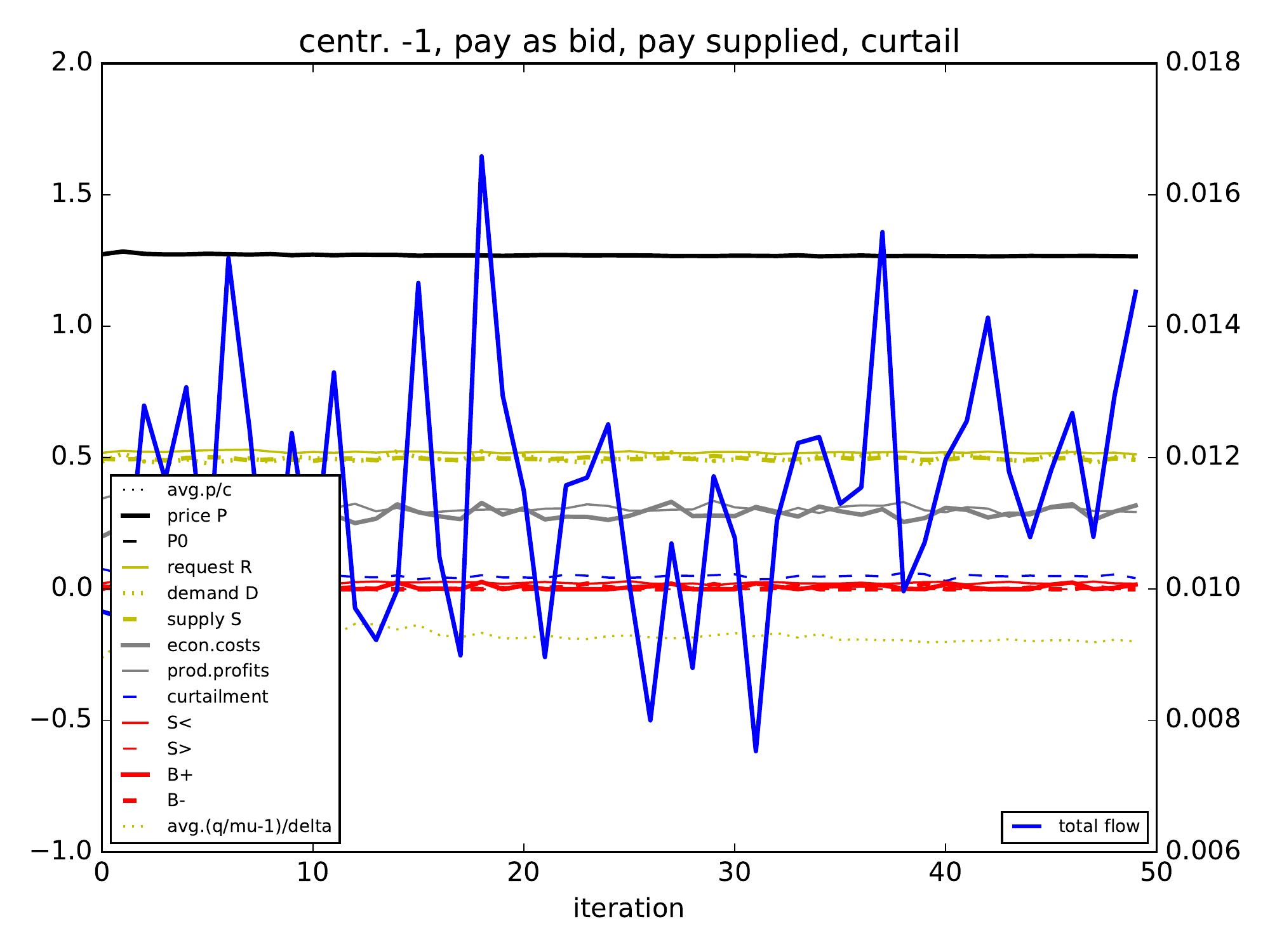} \\
	&
	\includegraphics[height=\hgt]{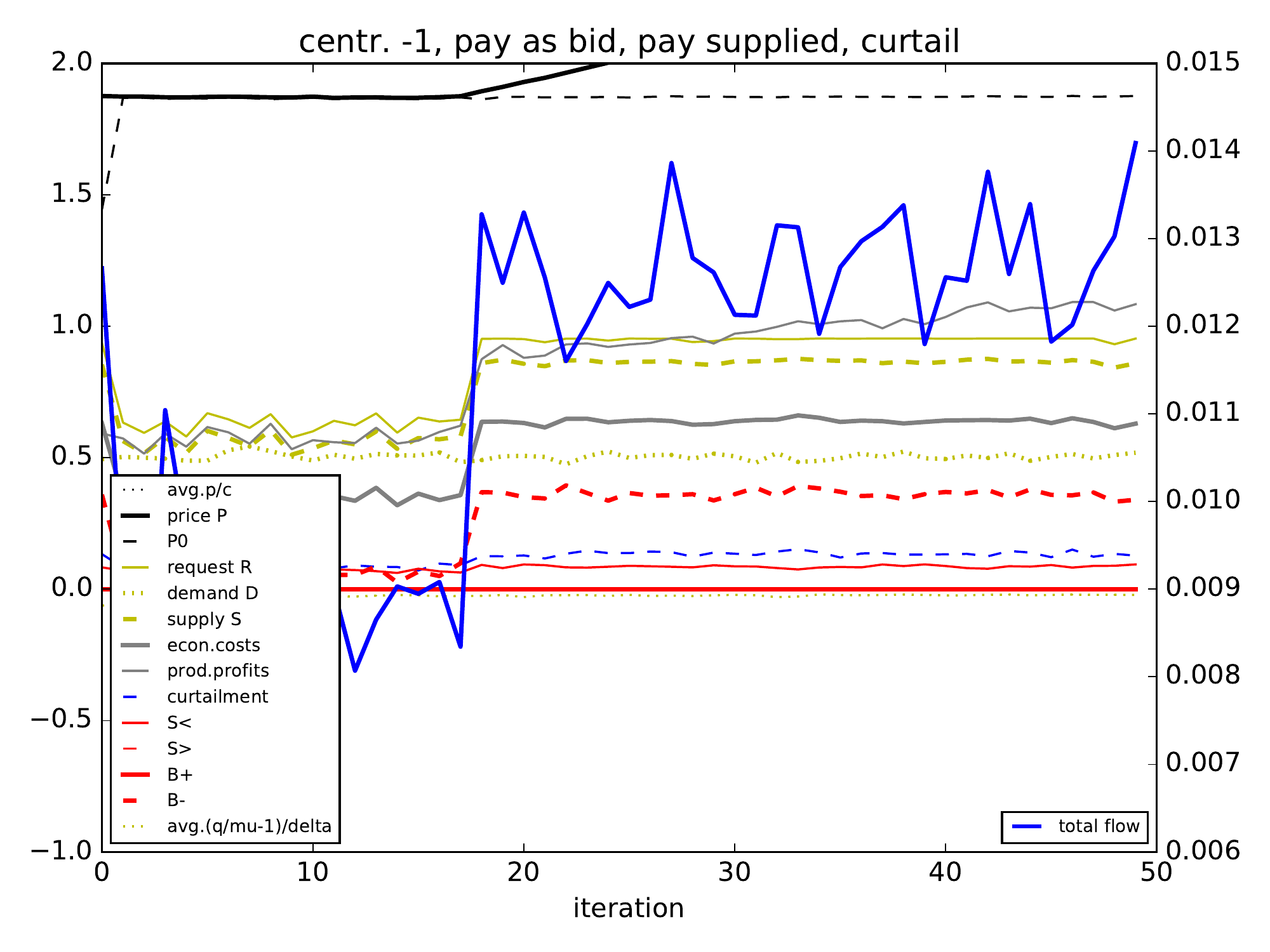} \\
	\includegraphics[height=\hgt]{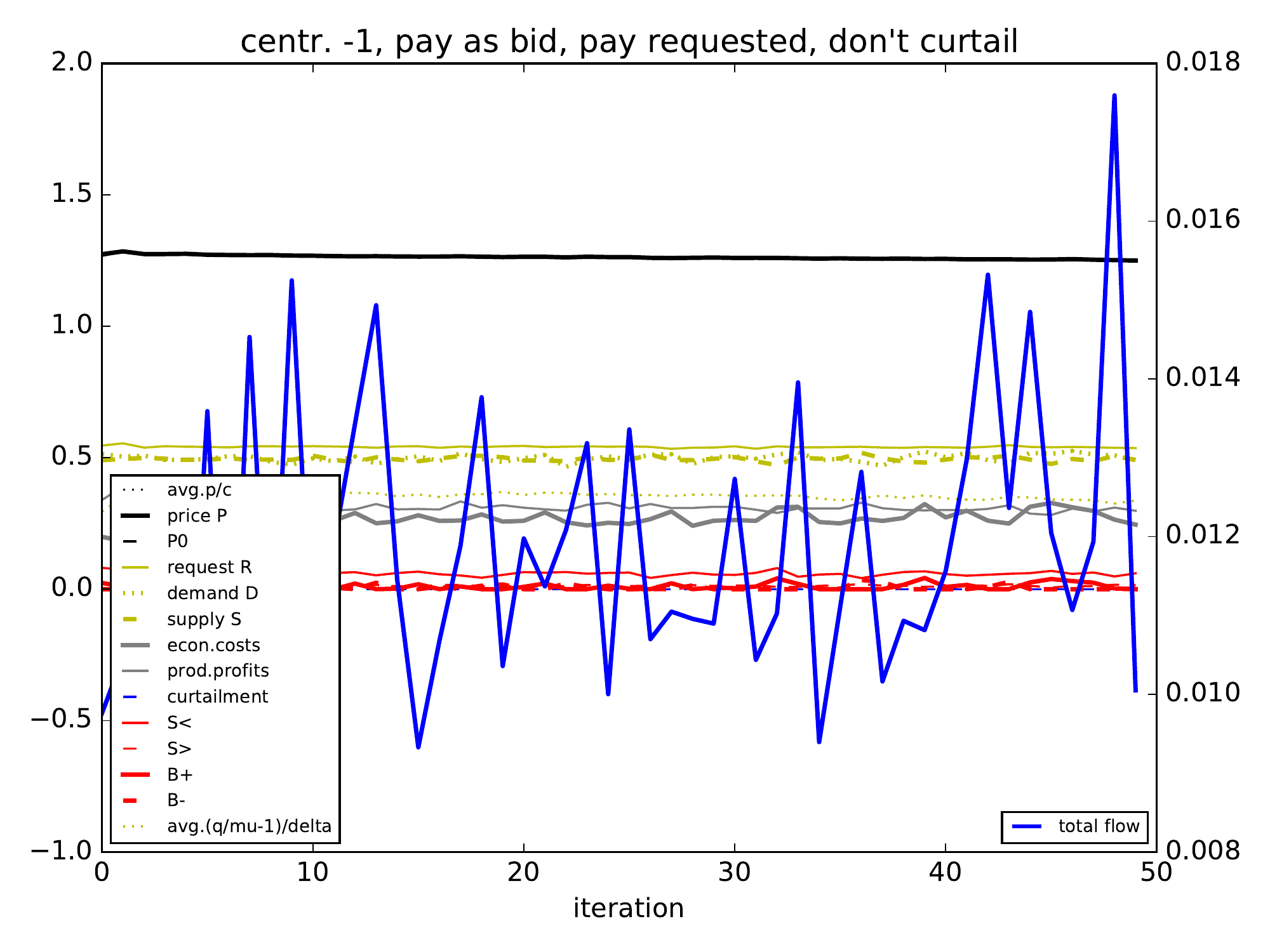} &
	\includegraphics[height=\hgt]{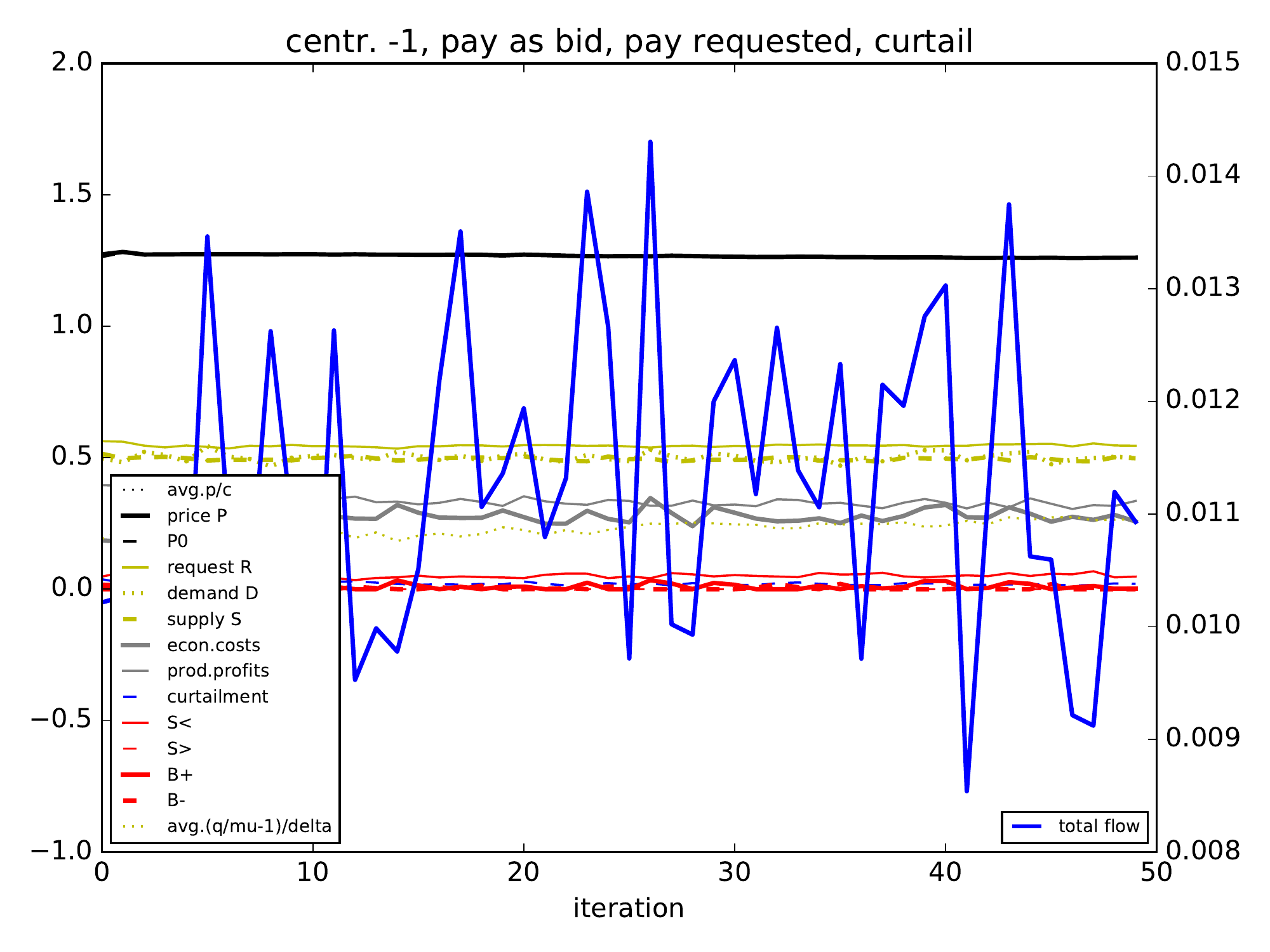} \\
\end{tabular}
\end{center}
\caption{\label{fig:trajectories-1}
Sample trajectories for all eight market design variants for the decentralized case $Z=-1$.
Again, in the ``pay as bid, pay supplied, curtail'' variant, 
an additional run with non-converging but increasing price and profits is shown.
}
\end{figure}

Our results are summarized in Table~\ref{tbl:results}, 
individual trajectories are shown in Figs.~\ref{fig:trajectories1}--\ref{fig:trajectories-1} for $Z\in{1,0,-1}$.

Overall, we find that in all but one marked design variant, 
bidding and price-setting behaviour converges  to a strategic equilibrium
already after a very small number of repetitions.
The only exception is the ``pay as bid, pay supplied, curtail'' (henceforth BSC) variant,
where for $Z\le 0$ in some runs, price bids start to increase linearly after a varying number of repetitions
(almost immediately in Fig.\,\ref{fig:trajectories0}, somewhat later in Fig.\,\ref{fig:trajectories-1}), 
causing the market price and hence the producer's profits to increase as well,
without affecting the total economic costs however (i.e., only causing an evergrowing transfer from consumers to producers).

\paragraph{Bid price.}
In general, bid prices are at least around twice as high as production costs,
being lowest under the ``uniform pricing, pay supplied, don't curtail'' design in the centralized case $Z=1$ ($p/c=1.89$ on average),
highest under the BSC design in the decentralized case $Z=-1$ ($p/c=14.62$ on average), 
which may be partially because of the increasing bid effect mentioned above,
but almost as high under the ``pay as bid, pay requested, don't curtail'' designs in the decentralized case ($p/c=12.16$ on average).
Bid prices are always lower under uniform pricing than under pay as bid,
since by bidding a lower price under uniforma pricing,
a producer will not risk getting a lower payment price (which equals $P$) but will only increase their chance of being selected,
hence the only incentive to not make a too low bid is to make sure that when selected, 
$P$ is large enough to cover production costs and expected penalties.

\paragraph{Bid quantity.}
For most designs, average bid quantities are below mean production capabilities:
producers bid conservative quantities since shortfall penalties are higher than excess supply penalties.
Only under ``pay as bid, pay requested'' for all values of $Z$
and under ``uniform pricing, pay requestes, don't curtail'' for $Z\le 0$, 
bid quantities are usually higher than mean production capabilities.
This is because in the ``pay requested'' designs, 
shortfall penalties are effectively lowered by the fact that the grid operator still pays for the (higher) bid quantity rather the (lower) actual supply.
The other two design choices (uniform pricing or pay as bid, and curtailment yes or no) were not found to have such a clear effect on bid quantities.

\paragraph{Market price and requested amount.}
Because of the lower bid prices, 
also market price is always lower under uniform pricing than under pay as bid, 
typically by a factor of around 1.5 to 2.
The extremes are
$P=2.37$ on average under BSC with $Z=-0.5$,
and $P=0.44$ on average under ``uniform pricing, pay supplied, don't curtail'' with $Z=-1$,
which is even below the average production costs, and consequently producers' profits are negative in this case,
making this a non-feasible design. 
The lowest price that makes production profitable is generally realized under ``uniform pricing, pay supplied, don't curtail'' for most values of $Z$,
resulting in $P\approx 1$, i.e., 1.5 to 3 times the mean production costs,
with profits still only being $\approx 0.15$ for $Z=1$ and $\approx 0.04$ for $Z=-1$ 
due to excess/shortfall penalties that grow with decreasing centralization.

The relationship between requested amount $R=\sum_{i:p_i\le P}q_i$ and actual demand $D$ differs widely between variants.
For all values of $Z$, they match closest under ``uniform pricing, pay requested, curtail'' and almost as well under ``uniform pricing, pay supplied, curtail''.
For $Z=-1$, they also match well under ``uniform pricing, pay requested, don't curtail''.
Under-requesting $D - R$ is largest under the ``pay supplied, don't curtail'' variants and still significant under the ``pay requested, don't curtail'' variants.
This is in part because if producers need not curtail excess production, 
the grid operator expects excess supply, knowing that bid (and thus requested) quantities 
are considerably (under ``pay supplied'') or at least moderately (under ``pay requested'') below mean production.
In contrast, in the ``pay as bid, curtail'' variants, $R$ is usually slightly larger than $D$.

\paragraph{Balancing needs.}
Interestingly, under the ``don't curtail'' variants, the grid operator's expectation of excess supply (i.e., positive $S - R$) 
turns out to be incorrect and consequently there is need for considerable upward balancing.
Except for the BSC design, upward balancing needs are generally significantly larger than downward balancing.
Under pay as bid, balancing needs are generally somewhat lower than under uniform pricing.

\paragraph{Total economic costs.}
Total economic costs, i.e., the sum of all production, curtailment, and balancing costs, but without penalties (which are only flows within the economy),
vary largely over the eight market design variants,
being largest under ``uniform pricing, pay supplied, don't curtail'' ($0.66$ on average) due to the large upward balancing needs,
and smallest (only half as high) under ``pay as bid, pay requested, curtail'' ($0.34$ on average).
This difference becomes larger for smaller centralization $Z$.
The second-best design w.r.t.\ total costs is ``uniform pricing, pay requested, curtail'' ($0.37$ on average).
In general, the choice between uniform pricing and pay as bid does not exhibit a consistent influence on total costs.
With the exception of ``uniform pricing, pay supplied, don't curtail'',
costs decrease with decreasing centralization $Z$, reflecting the lower production costs of VRES.

\paragraph{Producers' profits, consumers' costs.}
For an economic evaluation, producers' profits are the second most interesting indicator besides total economic costs.
They also vary largely over the eight market design variants,
being negative on average under ``uniform pricing, pay supplied, don't curtail'',
making this an economically infeasible design,
are slightly positive ($0.079$ on average) under ``uniform pricing, pay supplied, curtail'',
and are largest ($0.81$) under BSC.
Notably, under the overall cheapest design, ``pay as bid, pay requested, curtail'',
profits are large ($0.4$), 
while under the second-cheapest design, ``uniform pricing, pay requested, curtail'' (URC),
they are only a third as high ($0.13$).

Note that in our model, the grid operator's total costs equal total economic costs plus producers' profits,
and these can be interpreted as the consumers' total electricity costs.
Consequently, for ``pay as bid, pay requested, curtail'' these are about $0.34+0.4=0.74$ on average,
while for URC they are only about $0.37+0.13=0.5$.
When going from centralized to decentralized, this ratio increases from about $3/4$ to almost $1/2$.
Consumers' costs are even slightly smaller under ``uniform pricing, pay supplied, curtail'' 
($0.41+0.08=0.49$, with a strongly statistically significant two-sample $t$-test),
which is however considerably more expensive overall.
Hence consumers would prefer the latter design, producers would prefer BSC, 
a social planner not interested in welfare effects would prefer ``pay as bid, pay requested, curtail'',
and a social planner caring for welfare effects would probably prefer the compromise design URC.

\paragraph{Grid workload.}
As an overall indicator of the stress placed onto the grid, the potential needs for extending it, and the overall likelihood of line trippings,
we look at the total power flow summed over all grid lines, here called the ``grid workload'',
computed by standard power flow calculations on the underlying grid topology.
Overall, workload is smallest under the two ``uniform pricing, curtail'' variants,
with no statistically significant difference between the two,
and is highest and most volatile under BSC.
Except for the latter design, workload coefficient of variation is generally around $1/6$,
interestingly not varying much between the centralized and decentralized situations,
but suggesting that lines need a considerable amound of excess capacity to avoid tripping.

\section{Conclusion}

In view of the above results,
this study suggests that the best compromise between the eight possible designs studied here
might be given by the URC market design
in which producers are paid a uniform market price instead of their bid price,
are paid this price for the requested (i.e., the bid) quantity of electricity, 
regardless of whether their actual supply falls short of their bid quantity,
but have to pay a fixed penalty for each unit of electricity not supplied
and are not allowed to supply more than bid but have to curtail any excess production.
As seen above, this combination seems to lead to an almost optimal overall level of total economic costs of electricity production,
an almost optimal optimal level of consumers' costs,
a robust but moderate profitability for electricity producers,
a low level of balancing needs,
a moderate market price for electricity,
``honest'' bid quantities close to expected production capability,
and the smallest level of grid workload (i.e., higher reliability of supply).

All other studied designs seem to perform considerably worse than URC in one or several aspects.
The ``pay as bid'' variants have higher price bids, 
hence higher market prices and higher total consumers' costs,
less honest quantity bids,
higher workload,
and (except for ``pay requested, curtail'') higher total economic costs.
The ``pay supplied'' and ``don't curtail'' variants have
higher balancing needs, hence higher total economic costs,
and (except for ``uniform pricing, pay supplied, curtail'') higher workload.

In view of the rather conceptual nature of this study,
we recommend focussing in future research on
(i) simulating several coupled markets (e.g., day-ahead, intraday, balancing),
(ii) using real-world demand profiles and cost distributions for production, curtailment, and balancing,
(iii) distibnguishing different types of producers (traditional plants and utilities, small VRES producers, virtual power plants, etc.)
more explicitly than here, where we used a kind of continuum of producer types,
(iv) compare simulated bidding and price-setting behaviour over time with real-world data
to assess the amount of bounded rationality in real-world agents and improve the model.

\paragraph{Acknowledgements.}
The authors gratefully acknowledge the support of the German Federal Ministry of Education and Research (BMBF), CoNDyNet, FK. 03SF0472A,
and the European Regional Development Fund (ERDF), the German Federal Ministry of Education and Research and the Land Brandenburg 
for supporting this project by providing resources on the high performance computer system at the Potsdam Institute for Climate Impact Research.

\bibliographystyle{aer} 
\bibliography{bib}

\end{document}